\documentclass[twocolumn]{aastex63}

\usepackage{url}
\usepackage{amsmath}
\usepackage{amssymb}
\usepackage{float}
\usepackage{graphicx}
\usepackage{multirow}
\usepackage{soul}
\usepackage{rotating}
\usepackage{lineno}

\usepackage{xcolor}

\usepackage{xspace}

\usepackage{array}

\usepackage{hyperref}

\begin{document}

\title{Detecting Secular Perturbations in \emph{Kepler} Planetary Systems Using \\ Simultaneous Impact Parameter Variation Analysis (SIPVA)}


\author[0000-0003-2776-6092]{Zhixing Liu}
\affiliation{College of Letters and Science, University of California, Santa Barbara, Santa Barbara, CA 93106, USA}
\affiliation{Department of the Geophysical Sciences, University of Chicago, Chicago, IL 60637, USA}

\author[0000-0002-5668-243X]{Bonan Pu}  
\affiliation{Independent Researcher, New York, NY 10018, USA.}

\correspondingauthor{Zhixing Liu}
\email{zhixingliu@uchicago.edu}

\begin{abstract}
Secular impact-parameter variations encode dynamical perturbations and can help reveal unseen companions or constrain planetary masses when combined with dynamical interpretation. Existing methods either fit transits independently or jointly model the light curves and orbital dynamics.
We introduce \emph{Simultaneous Impact Parameter Variation Analysis} (SIPVA), which avoids the N-body integrations used in many dynamical or photodynamical fitting approaches. SIPVA directly incorporates a linear time-dependent model for impact parameters into the Markov Chain Monte Carlo (MCMC) framework by fitting all transits simultaneously.
We evaluate SIPVA and the transit-by-transit implementation on artificial systems grouped by the fixed-model injection-strength label $L\equiv\sqrt{\mathrm{LLR}_{\rm theory}}$. In these controlled injections, SIPVA outperforms the particular transit-by-transit implementation examined here in both detection and recovery performance. Applied to 16 \emph{Kepler} planets previously identified as exhibiting significant transit-duration variations, SIPVA detects impact-parameter trends in 9 planets, compared with 5 using independent transit fits.
SIPVA provides a joint light-curve framework for measuring secular impact-parameter trends without specifying a dynamical architecture or performing N-body integrations.
\end{abstract}


\section{Introduction} \label{sec:intro}

Accurately estimating the mass and eccentricity of exoplanets remains a significant challenge in detecting and characterizing \emph{Kepler} multi-planet systems. One common approach involves reproducing observational data through N-body simulations to determine planetary masses \citep{1991AJ....102.1528W, 1999MNRAS.304..793C, 2011Sci...333.1602D, 2012Sci...337..556C}, which has contributed multiple measurements for \emph{Kepler} systems. Despite advancements in numerical integrators \citep{2012A&A...537A.128R, Rein_2014, 2015MNRAS.452..376R, Rein_2019_new, Rein_2019, Javaheri_2023}  and increased computational capabilities, this method is limited by the \emph{Kepler} mission's approximately four-year observational window, which is insufficient to capture the frequent yet subtle orbital parameter changes necessary for precise mass determination using N-body techniques.

Transit Timing Variation (TTV) analysis is an effective method for estimating exoplanet masses \citep{2005MNRAS.359..567A, Grimm2018, Dai2023, Masuda2024}. Several tools, including \texttt{ttvim} \citep{Nesvorn__2010}, \texttt{TTVFast} \citep{Deck_2014}, \texttt{TTVFaster} \citep{2016ascl.soft04012A}, \texttt{PhoDyMM} \citep{2022BAAS...54e.360J}, and \texttt{PyDynamicaLC} \citep{2021ApJ...908..114Y}, facilitate efficient TTV analyses. {In addition to constraining planetary masses, studies have analyzed the role of eccentricity by examining the fundamental TTV modes (the long-period sinusoidal variations near a mean-motion resonance) \citep{2012ApJ...761..122L, 2016ApJ...818..177A}, and extending them to higher orders \citep{2017AJ....154....5H}.} However, TTV analysis is challenged by mass and eccentricity degeneracies \citep{2012ApJ...761..122L}, complicating the precise determination of each individual parameter. {Moreover, planets with shorter orbital periods have shorter transit durations, resulting in fewer data points within each transit. This reduced temporal resolution makes detecting and analyzing subtle TTV signals more difficult}.

Transit Duration Variation (TDV) analysis, especially when combined with TTV, can help overcome the mass-eccentricity degeneracy \citep{Agol_2018}. Recent studies have actively explored using TDVs to constrain exoplanetary parameters \citep{Shahaf_2021, Judkovsky2020}. Physical evidence for secular changes in TDVs driven by variations in impact parameters resulting from mutual inclinations has been presented by \citet{2021AJ....162..166M}. Impact-parameter evolution is not restricted to observed multiplanet systems: apparently single-planet systems can also exhibit it through stellar-quadrupole-driven orbital precession, with perturbations from an undetected companion providing another possible cause \citep{Miralda_Escude_2002, Szab__2012}. Consequently, analyzing impact parameter variations (TbVs) has emerged as an alternative approach. Using TbVs, \citet{Judkovsky_2022} analyzed 54 systems with 140 planets, achieving mass detections of more than $3\sigma$ for 102 planets, including 43 with masses below $5\,M_{\oplus}$. Their subsequent research, \citet{judkovsky2024kepler}, estimates orbital parameters for 101 planets across 23 systems, with mass significances better than $3\sigma$ for 95 planets, including 46 without prior constraints.

Recovering and precisely estimating the magnitude and direction of TDVs or TbVs is crucial before using them to constrain planetary parameters. Traditionally, two methodologies have been employed: the \emph{individual fit}, which fits each transit independently to analyze TDVs \citep{Holczer_2016}, and the \emph{dynamical fit}, which simulates planetary dynamics to match transit light curves \citep{Judkovsky_2022, Judkovsky_2022b, judkovsky2024kepler, Langford2025}.

The \emph{dynamical fit} requires assumptions about the system architecture---including the number of planets, their masses, and their orbital elements---to initialize N-body integrations, as in analyses of Kepler-79 and Kepler-11 \citep{judkovsky2024kepler}. SIPVA instead models each planet independently of its neighbors, fitting all of its transits simultaneously with a linear time-dependent impact parameter in an MCMC framework. By avoiding full dynamical modeling, SIPVA reduces computational cost and reliance on prior architectural assumptions. In the controlled synthetic tests presented here, SIPVA achieved better detection and recovery than the particular transit-by-transit implementation examined in this work.


Sections~2--3 develop the transit geometry and define the theoretical model-separation statistic. Section~4 presents the individual-fit and SIPVA methods and their synthetic tests, Section~5 applies both methods to 16 \emph{Kepler} candidates, and Section~6 summarizes the results.

\section{Transit Geometry and the Observability of Impact Parameter Variations} \label{sec:transit_geometry}
\subsection{Forward Transit Model and Parameterization}
We describe each transit light curve using the non-impact-parameter model vector
\begin{equation}\label{eq:theta_circular}
\vec{\Theta} =\left\langle t_e, P, p, \rho_{\rm eff}, q_1, q_2\right\rangle .
\end{equation}
Here $\vec{\Theta}$ denotes only the non-impact-parameter model arguments; the complete forward-model parameter set is $(\vec{\Theta},b)$, where $b$ is the sky-projected chord--center separation in stellar-radius units; $b=0$ corresponds to a central transit, and larger $b$ places the transit chord closer to the stellar limb. Within $\vec{\Theta}$, $t_e$ is the mid-transit time, $P$ is the orbital period, $p=R_p/R_{\star}$ is the planet-to-star radius ratio, and $\rho_{\rm eff}$ is the fitted stellar-density coordinate defined below.

Throughout this section and in Sections~\ref{sec:LLR} and \ref{sec:Methodology}, we set the eccentricity to zero, $e=0$, while parameterizing the inclination dependence through $b$. For quadratic limb darkening, we adopt the reparameterization of \citet{Kipping2013},
\begin{equation}
u_1=2 \sqrt{q_1} q_2, \quad u_2=\sqrt{q_1}\left(1-2 q_2\right),
\end{equation}
which maps the physically allowed region in $\left(u_1, u_2\right)$ onto the unit square in $\left(q_1, q_2\right)$. We evaluate the transit light curve using the analytic quadratic limb-darkened formalism of \citet{Mandel_2002}.

The transit shape depends on the scaled semimajor axis $a_R\equiv a/R_\star$. Retaining both the stellar and planetary masses in Kepler's third law,
\begin{equation}
\begin{aligned}
\left(\frac{a}{R_\star}\right)^3
&=\frac{G P^2(M_\star+M_p)}{4\pi^2 R_\star^3} \\
&=\frac{G P^2}{3\pi}\left(\rho_\star+p^3\rho_p\right)
\equiv\frac{G P^2}{3\pi}\rho_{\rm eff},
\end{aligned}
\end{equation}
where $\rho_p$ is the planet's mean density and $\rho_{\rm eff}\equiv\rho_\star+p^3\rho_p$. Among the 16 planets analyzed in Section~\ref{sec:real_kepler}, the subset with reliable mass measurements has $p^3\rho_p/\rho_\star\sim10^{-5}$--$10^{-3}$, well below the adopted density-prior widths and typical $3\%$ posterior uncertainties. We therefore retain priors centered on the catalog $\rho_\star$ and interpret the fitted density as $\rho_{\rm eff}$; including the planetary term would not materially change the inferred transit parameters.
In the general Keplerian case, the sky-projected separation depends on $a_R$, $e$, $\omega$, and the viewing geometry. For the circular synthetic systems considered here, however, it can be written directly in terms of the fitted impact parameter $b$. Here $z(t)$ is the sky-projected separation between the planet and stellar centers, expressed in units of $R_{\star}$.
\begin{equation} \label{eq:z_squared_t}
  z^2(t)=b^2+\left[\left(\frac{a}{R_{\star}}\right)^2-b^2\right] \sin ^2\left[\frac{2 \pi\left(t-t_e\right)}{P}\right] .
\end{equation}
Near central transits, the impact parameter becomes poorly constrained because standard error propagation gives $\mathrm{SE}_b \simeq \mathrm{SE}_{b^2}/(2b)$, where $\mathrm{SE}_b$ and $\mathrm{SE}_{b^2}$ are the standard errors of $b$ and $b^2$, respectively. Therefore, when $b$ is small, even a modest uncertainty in $b^2$ can produce an uncertainty in $b$ that is comparable to, or larger than, $b$ itself.

With these definitions, the quadratic limb-darkened transit model is
\begin{equation}
\begin{aligned}
\hat{F}(t ; \vec{\Theta}, b)
&=1-\frac{1}{C_{\mathrm{LD}}}\Bigl(\left[1-2 \sqrt{q_1}\left(1-q_2\right)\right] \lambda^e \\
&\qquad +2 \sqrt{q_1}\left(1-q_2\right)\left[\lambda^d+\frac{2}{3} H(p-z)\right] \\
&\qquad +\sqrt{q_1}\left(1-2 q_2\right) \eta^d\Bigr),
\end{aligned}
\end{equation}
where $C_{\mathrm{LD}}=1-\frac{\sqrt{q_1}\left(1+2 q_2\right)}{6}$ is the limb-darkening normalization, $H$ is the Heaviside step function, and $\lambda^e$, $\lambda^d$, and $\eta^d$ are the occultation functions of \cite{Mandel_2002}, evaluated at $(p, z(t))$.
Because the model depends on $b$ only through $z^2(t)$, the transit light curve is even in $b$, i.e. $\hat{F}(t ; \vec{\Theta}, b)=\hat{F}(t ; \vec{\Theta},-b)$, and therefore directly constrains $b^2$, not the signed $b$.
Throughout this work, $b$ denotes the non-negative sky-projected separation. Thus, $\dot{b}<0$ means that the transit chord is moving toward the stellar center. The linear model assumes that $b(t)$ does not cross $b=0$ over the fitted baseline.
\subsection{Transit Duration and Its Time Derivative}
Because $b$ is often poorly constrained, previous studies have used transit-duration variations (TDVs) as an indirect probe of secular orbital evolution \citep{Miralda_Escude_2002, Boley_2020, Judkovsky2020, Shahaf_2021, Szab__2012, 2021AJ....162..166M}.
The first and fourth contact points occur when the projected planet--star center separation satisfies $z(t)=1+p$. For a circular orbit, these endpoints are the two symmetric times about $t_e$, $t=t_e \pm \frac{T_{\mathrm{dur}}}{2}$.
Substituting this into Equation~(\ref{eq:z_squared_t}) gives,
\begin{equation}
(1+p)^2=b^2+\left[\left(\frac{a}{R_{\star}}\right)^2-b^2\right] \sin ^2\left(\frac{\pi T_{\mathrm{dur}}}{P}\right)
\end{equation}
and solving for $T_{\mathrm{dur}}$ yields
\begin{equation}
T_{\mathrm{dur}}=\frac{P}{\pi} \arcsin \left[\left(\frac{(1+p)^2-b^2}{\left(a / R_{\star}\right)^2-b^2}\right)^{1 / 2}\right] .
\end{equation}
Let $\tau$ denote the ingress or egress duration. Replacing $T_{\text {dur }}$ by $T_{\text {dur }}-2 \tau$ and $1+p$ by $1-p$ then gives the analogous relation for the interior part of the transit.
For the compact systems considered here, $R_{\star} / a \ll 1$, so the argument of the arcsine is small and $\arcsin x \approx x$. Using $\left(a / R_{\star}\right)^2 \gg b^2$ and $p \ll 1$ the total duration therefore becomes
\begin{equation}
T_{\mathrm{dur}} \approx \frac{P}{\pi} \frac{R_{\star}}{a} \sqrt{1-b^2} .
\end{equation}
Expanding the contact relations to first order in $p$ also gives the ingress/egress duration
\begin{equation}
\tau \approx \frac{P}{\pi} \frac{R_{\star}}{a} \frac{p}{\sqrt{1-b^2}} .
\end{equation}
If the transit depth is $\delta \approx p^2$, then these two observables imply
\begin{equation}
b^2 \approx 1-\sqrt{\delta} \, \frac{T_{\mathrm{dur}}}{\tau},
\qquad
\frac{R_{\star}}{a} \approx \frac{\pi}{P}\left(\frac{T_{\mathrm{dur}}\tau}{\sqrt{\delta}}\right)^{1 / 2},
\end{equation}
Thus, in the circular limit, the observable set $(\delta, T_{\mathrm{dur}}, \tau, P, t_e)$ determines $(p, b^2, a / R_{\star})$ \citep{Seager_2003}. In practice, the timing parameters $P$ and $t_e$ are typically constrained much more precisely than the shape observables $T_{\mathrm{dur}}, \tau$, and $\delta$, so the latter dominate the uncertainty propagation relevant to $b$.
Let $v_{\perp,\mathrm{mid}}$ denote the sky-projected orbital velocity across the transit chord at mid-transit. For the circular synthetic systems considered here,
$v_{\perp,\mathrm{mid}}=2\pi a/P$,
so the same expression may be written as
\begin{equation}
T_{\mathrm{dur}} \approx \frac{2 R_{\star} \sqrt{1-b^2}}{v_{\perp,\mathrm{mid}}} .
\end{equation}
Taking the derivative of the above expression gives
\begin{equation}\label{eq:tdv_derivative}
\dot{T}_{\mathrm{dur}}=-T_{\mathrm{dur}}\left[\dot{b} \frac{b}{1-b^2}+\frac{\dot{v}_{\perp,\mathrm{mid}}}{v_{\perp,\mathrm{mid}}}\right] .
\end{equation}

This is the TDV relation used in Eq.~(4) of \citet{2021AJ....162..166M}. In general, $\dot{T}_{\mathrm{dur}}$ depends on both $\dot{b}$ and $\dot{v}_{\perp,\mathrm{mid}}$, and the projected transit velocity may evolve in eccentric systems through orbital-element evolution, including apsidal precession, which changes the orbital phase and speed sampled at mid-transit. TDVs therefore trace changes in transit geometry but are not generally a one-to-one proxy for $\dot b$. For the circular synthetic systems considered here, $v_{\perp,\mathrm{mid}}=2\pi a/P$ is constant, so $\dot v_{\perp,\mathrm{mid}}=0$ and $\dot{T}_{\mathrm{dur}}=-T_{\mathrm{dur}} \dot{b} \frac{b}{1-b^2}$.

\section{Fixed-Model Theoretical Separation} \label{sec:LLR}

Equation~(\ref{eq:tdv_derivative}) connects impact-parameter evolution to transit duration, but the same evolution also changes the full light-curve profile through $z(t)$. Because SIPVA fits the flux time series directly, we define a fixed-model measure of the expected separation between a constant-$b$ model $M_0$ and a linear-$b(t)$ model $M_1$ using a noise-averaged log-likelihood ratio.

\begin{equation}\label{eq:LLR_definition}
    \mathrm{LLR}=2\left(\ln \mathcal{L}_1-\ln \mathcal{L}_0\right),
\end{equation}
where $\mathcal{L}_1$ and $\mathcal{L}_0$ are the likelihoods of the data under models $M_1$ and $M_0$, respectively. The two parameter vectors are fixed: $M_1$ uses the injected linear $b(t)$ trajectory, while $M_0$ holds $b=b_0$ and uses the same values of all other parameters. In particular, $M_0$ is not optimized over a constant impact parameter or profiled over any other parameter. For data generated under the fixed $M_1$ vector with i.i.d. Gaussian flux errors, the expectation of Equation~(\ref{eq:LLR_definition}) is
\begin{equation} \label{eq:LLR_square}
\begin{aligned}
\mathrm{LLR}_{\rm theory}
&\equiv
\mathbb{E}_{M_1}\!\left[\mathrm{LLR}\right] \\
&=
\frac{1}{\sigma_{\rm phot}^{2}}
\sum_{j=1}^{N_T}
\sum_{i=1}^{n_j}
\left[
\widehat F_{M_1}\left(t_{ij}\right)
-
\widehat F_{M_0}\left(t_{ij}\right)
\right]^2 .
\end{aligned}
\end{equation}
Here, $\widehat F_{M_1}$ and $\widehat F_{M_0}$ are the noiseless mean fluxes predicted by the two fixed parameter vectors at the same sampled times. Thus, $\mathrm{LLR}_{\rm theory}$ measures their expected separation for an assumed variance $\sigma_{\rm phot}^2$; it is not a generalized or profile likelihood-ratio statistic comparing best-fitting models. Here $t_{ij}$ is the time of data point $i$ in transit $j$, $n_j$ is the number of sampled points in that transit, and $N_T$ is the number of transits. The dependence on the projected separation $z(t_{ij})$ is implicit in the transit model $\widehat F(t_{ij})$, and $\sigma_{\rm phot}$ is the assumed per-cadence flux uncertainty. Equation~(\ref{eq:LLR_square}) can be evaluated for any specified set of observation times and assumed per-cadence uncertainty.

{The adopted value of $\sigma_{\rm phot}$ fixes the normalization of $\mathrm{LLR}_{\rm theory}$. The synthetic comparison should therefore be interpreted conditional on the adopted i.i.d. Gaussian-noise model.}

In the recovery analysis, $L\equiv\sqrt{\mathrm{LLR}_{\rm theory}}$, so $L=5,10,20,30,$ and $50$ correspond to $\mathrm{LLR}_{\rm theory}=25,100,400,900,$ and $2500$, respectively. We use $L$ only as a fixed-model injection-strength label for constructing the synthetic categories. Detection uses the method-specific standardized statistics defined in Section~\ref{sec:Methodology}.

\section{Methodology} \label{sec:Methodology}

We present two MCMC-based methods for detecting impact parameter variations: the \emph{individual fit} and SIPVA. Section \ref{sec:Generating_Artificial_transit} describes Monte Carlo simulations used to generate artificial light curves at fixed target $L$ values. Section \ref{sec:individual_fit_1} outlines the \emph{individual fit} approach, where each transit is fit independently, while Section \ref{sec:Group_fit} introduces SIPVA, which incorporates a linear time-dependent impact parameter into the MCMC framework. Section \ref{sec:results} then compares the SIPVA impact-parameter fit, the transit-by-transit impact-parameter regression, and the transit-by-transit TDV regression by evaluating recovery rates and ROC curve performance.

\subsection{Generating Artificial Systems at Fixed-Model Injection-Strength Categories} \label{sec:Generating_Artificial_transit}
We generated 50 accepted synthetic systems in each of the five non-null
categories $L=5,10,20,30,$ and $50$, corresponding to
$\mathrm{LLR}_{\rm theory}=25,100,400,900,$ and $2500$, respectively. For
each candidate, we drew $P$ log-uniformly on $[13.432,44.773]$~d,
$b_0\sim\mathcal{U}(0.07,0.5)$, $p$ log-uniformly on $[0.01,0.06]$, and
$a/R_\star$ log-uniformly on $[30,100]$. The number of retained transits was
$N_T=\left\lfloor1460\times0.92/P\right\rfloor\in[30,100]$, with centers
$t_{c,j}=jP$ for $j=0,1,\ldots,N_T-1$. The injected density
coordinate was derived as
$\rho_{{\rm eff},0}=3\pi(a/R_\star)^3/(GP^2)$.

Each transit was sampled at one-minute cadence in a fixed five-hour window centered on $t_{c, j}$, with 300 cadences and no gaps within the retained window. Depending on the $L$ category, $T_{14}>5 \mathrm{hr}$ for approximately 12--22\% of accepted systems, so these segments do not contain the complete transit or an out-of-transit baseline.

The constant-impact-parameter model $M_0$ keeps $b=b_0$ at every transit, whereas the injected model $M_1$ evolves as $b_j=b_0+\dot{b} P j / 365$; neither fixed parameter vector is optimized during this target-matching calculation. For each non-null draw, we first evaluated the fixed-model label $L$ at $\dot b=0.01$ and $0.03 \mathrm{yr}^{-1}$. If the target $L_{\text {target }}$ lay between these two values, Brent's method was used to solve $L(\dot{b})=L_{\text {target }}$ from the noiseless $M_1-M_0$ flux difference; otherwise, the system was discarded and redrawn. We also rejected systems for which the final transit would no longer satisfy $b_{N_T-1}<1+p$. Because the search interval permits only positive trends, all injected $\dot{b}$ values lie between 0.01 and $0.03 \mathrm{yr}^{-1}$. This selection changes the accepted population across $L$ categories: low- $L$ bins reject systems whose signal is already too strong at $\dot{b}=0.01$, whereas high- $L$ bins reject systems that cannot reach the target even at $\dot{b}=0.03$.

Only after target matching did we add independent Gaussian noise with
$\sigma_{\rm phot}=2\times10^{-4}$ to each normalized-flux sample. This noise
level corresponds to 200 ppm per \emph{Kepler} short-cadence exposure,
comparable to the near-Poisson-limited precision achieved by \emph{Kepler} at
11th magnitude \citep{Gilliland_2011}. The matched null set reused the system
parameters of the 50 accepted $L=10$ systems, set $\dot b=0$, and used
independent noise realizations, giving 300 systems in total. The synthetic
injections use fixed limb darkening $(u_{1,0},u_{2,0})=(0.43,0.38)$, equivalent
to $(q_{1,0},q_{2,0})=(0.656,0.265)$ under the \citet{Kipping2013}
reparameterization, whereas the recovery fits sample $q_1$ and $q_2$.
Circular orbits were assumed throughout.

{After target matching, the noisy fluxes used in the recovery analysis are}
\begin{equation} \label{eq:synthetic_flux_noise}
F_{ij}=\widehat F_{M_1}\left(t_{ij}\right)+\epsilon_{ij},
\qquad
\epsilon_{ij}\overset{\mathrm{i.i.d.}}{\sim}
\mathcal{N}\!\left(0, \sigma_{\rm phot}^{2}\right) .
\end{equation}
{Here, ``independent'' means independent across sampled data points.}

\subsection{Initial Transit Fitting Approach} \label{sec:individual_fit_1}
{In the initial transit-by-transit analysis, the complete sampled parameter set for the $j$th transit is $(\vec{\Theta}_j,b_j)$, where the non-impact-parameter vector corresponding to Equation~(\ref{eq:theta_circular}) is
\[
\vec{\Theta}_j
=\left\langle t_{{\rm e},j},P,p,\rho_{\rm eff},q_1,q_2\right\rangle,
\]
and $b_j$ is the impact parameter. Thus, each transit fit has seven free parameters. All seven are re-estimated for each transit, although the transit index is suppressed for the components of $\vec{\Theta}_j$ other than $t_{{\rm e},j}$. For the synthetic individual fits in this section, the known injected scale $\sigma_{\rm phot}$ is used in the likelihood.}

{We implement these independent fits with MCMC, using \texttt{PyTransit} \citep{Parviainen_2015} for the transit model and \texttt{emcee} \citep{Foreman_Mackey_2013} for sampling.}

{Following the Kepler DR25 analysis \citep{Thompson_2018}, parameters that are weakly constrained by a single transit (e.g. $P$, $\rho_{\rm eff}$, and limb darkening) are anchored by informative priors, while $b_j$ and $t_{{\rm e},j}$ are allowed to vary more freely on a per-transit basis. For each transit, we assign Normal priors to $t_{{\rm e},j}$, $P$, $p$, $\rho_{\rm eff}$, and $b_j$, and Uniform priors to $q_1$ and $q_2$. The prior on $t_{{\rm e},j}$ is centered on the median flux minimum with $\Delta_{t_e}^{\mathrm{pr}}=0.5\,\mathrm{d}$ (weakly informative). To mimic imperfect literature constraints, the Normal-prior means for $P$, $p$, $b$, and $\rho_{\rm eff}$ are offset from the simulation truth by $3 \times 10^{-5} P$, $3 \%$, $\pm 0.05$, and $10 \%$, respectively; the corresponding prior widths are $10^{-5} P$, $\Delta_p^{\mathrm{pr}}=p/50$ (a $2\%$ fractional width), $0.2$, and $\rho_{\rm eff} / 10$. For the synthetic recovery tests, we use $q_1,q_2 \sim \mathcal{U}(0,1)$; thus $q_1$ and $q_2$ are sampled rather than fixed to the injected values. These prior distributions are reused for every transit, but the sampled parameters are not shared across epochs.}

The logarithmic likelihood function for the $j$th synthetic transit, $\ell_j(\vec{\Theta}_j,b_j \mid F_j,\sigma_{\rm phot})$, is formulated from the observed fluxes $F_{ij}$ and the model flux $\hat{F}(t_{ij};\vec{\Theta}_j,b_j)$ as
\begin{equation} \label{eq:normalized_log-likelihood}
\begin{aligned}
\ell_j(\vec{\Theta}_j,b_j \mid F_j,\sigma_{\rm phot})
&= -\frac{1}{2}\sum_{i=1}^{n_j}
\left[
\frac{F_{ij}-\hat{F}\!\left(t_{ij};\vec{\Theta}_j,b_j\right)}{\sigma_{\rm phot}}
\right]^2 \\
&\quad -\frac{n_j}{2}\log\!\left(2\pi\sigma_{\rm phot}^2\right).
\end{aligned}
\end{equation}
For any normally distributed parameter $x$, we denote its prior density by $\mathcal{N}\!\left(x;\mu_x,\left(\Delta_x^{\mathrm{pr}}\right)^2\right)$, where $\Delta_x^{\mathrm{pr}}$ is the adopted prior width.
The prior distribution is then expressed as the sum of contributions from parameters with Normal priors and those with Uniform priors,
\begin{equation}\label{eq:prior_updated}
\begin{aligned}
\log \pi(\vec{\Theta}_j,b_j)
&= \sum_{i\in\mathcal{N}} \log \mathcal{N}\!\left(
\Theta_{j,i};\mu_{\Theta_i},(\Delta_{\Theta_i}^{\mathrm{pr}})^2\right) \\
&\quad + \log \mathcal{N}\!\left(
b_j;\mu_{b_j},(\Delta_{b_j}^{\mathrm{pr}})^2\right) \\
&\quad + \sum_{k\in\mathcal{U}}\log \mathrm{Unif}(\Theta_{j,k};a_k,b_k).
\end{aligned}
\end{equation}
The logarithmic posterior distribution of the transit-specific parameters $\vec{\Theta}_j$ and $b_j$, based on the observed flux $F_j$, combines the prior terms with the likelihood of observing $F_j$ given the model:
\begin{equation}\label{eq:posterior_updated}
\begin{aligned}
\log P(\vec{\Theta}_j,b_j \mid F_j,\sigma_{\rm phot})
&=
\ell_j(\vec{\Theta}_j,b_j \mid F_j,\sigma_{\rm phot})
\\
&\quad+\log \pi(\vec{\Theta}_j,b_j).
\end{aligned}
\end{equation}
{Each transit fit samples the six components of $\vec{\Theta}_j$ together with $b_j$, for seven free parameters in total. We used an ensemble MCMC with 100 walkers. Prior to sampling, we performed a differential‐evolution global optimization (750 iterations) with a population of 100 candidate solutions to initialize the ensemble. We then ran three repeats of 1500 iterations each, storing every 20th step (thinning = 20), and discarded the first 25\% of samples in each chain as burn‐in. Convergence was checked via trace plots and posterior histograms, which showed stationary chains and unimodal posteriors for all parameters.}

Each transit is summarized by the posterior median $q_{50,j}$ and an effective symmetric 68\% posterior uncertainty $\Delta b_j$, defined as the larger of the 16th--50th percentile half-width $q_{50,j}-q_{16,j}$ and the 50th--84th percentile half-width $q_{84,j}-q_{50,j}$. The per-transit posterior medians are converted to a long-term trend with a weighted linear regression against the fixed model-minimum centers $t_{c,j}$ determined after the individual fits,
\begin{equation}
q_{50,j} = b_{\rm ref}
+ m_b^{\rm IF}\left(t_{c,j}-t_{\rm ref}\right)
+ \epsilon_j,
\end{equation}
using weights $w_j=(\Delta b_j)^{-2}$. The fitted slope $m_b^{\rm IF}$ serves as the transit-by-transit baseline used here for comparison with SIPVA.

We denote the formal weighted-regression standard error of the impact-parameter slope by $\sigma_{m_b^{\mathrm{IF}}}$ and define
\begin{equation}
Z_b^{\mathrm{IF}}
\equiv
\frac{m_b^{\mathrm{IF}}}{\sigma_{m_b^{\mathrm{IF}}}}.
\end{equation}
This is a signed, dimensionless slope-to-uncertainty ratio.

Transit durations are computed from the posterior samples of each independently fitted epoch. Let $T_{14,50,j}$ denote the posterior-median total duration for epoch $j$, evaluated from its finite, transiting posterior samples. We define its effective symmetric 68\% posterior uncertainty as
\begin{equation}
\Delta T_{14,j}
=
\max\left[
T_{14,50,j}-T_{14,16,j},
T_{14,84,j}-T_{14,50,j}
\right].
\end{equation}
Thus, $\Delta T_{14,j}$ is the posterior uncertainty on the duration of one transit. Epochs for which more than half of the posterior samples yield non-finite durations because the geometry is non-transiting are excluded from the TDV regression. Otherwise, the duration percentiles are calculated from the finite, transiting samples. The regression is performed only when at least five valid epochs remain; all 300 systems in the reported synthetic experiment meet this requirement.

We estimate the long-term duration trend using the weighted linear regression
\begin{equation}
T_{14,50,j}
=
\alpha_T
+ m_T^{\mathrm{IF}}t_{c,j}
+ \epsilon_{T,j},
\end{equation}
where $T_{14,50,j}$ is expressed in minutes, $t_{c,j}$ is expressed in days, and the weights are $w_j=(\Delta T_{14,j})^{-2}$. The fitted duration slope $m_T^{\mathrm{IF}}$ therefore has units of minutes per day. We denote its formal weighted-regression standard error by $\sigma_{m_T^{\mathrm{IF}}}$. The duration uncertainties are treated as absolute in calculating the regression covariance, so the slope uncertainty is not rescaled by the reduced chi-squared.

We define the signed, dimensionless TDV statistic as
\begin{equation}
Z_T^{\mathrm{IF}}
\equiv
\frac{m_T^{\mathrm{IF}}}{\sigma_{m_T^{\mathrm{IF}}}}.
\end{equation}
Here, $\sigma_{m_T^{\mathrm{IF}}}$ is the regression standard error of the long-term duration slope, distinct from the per-transit posterior uncertainty $\Delta T_{14,j}$. The statistic $Z_T^{\mathrm{IF}}$ is a standardized regression slope, not a posterior parameter or a Student's t-statistic. For the circular synthetic systems, the projected mid-transit velocity is constant, and Equation~(\ref{eq:tdv_derivative}) implies that the positive injected impact-parameter rates, together with $b>0$, produce negative duration slopes. The expected TDV signal direction is therefore $Z_T^{\mathrm{IF}}<0$, and a synthetic TDV recovery is declared when $Z_T^{\mathrm{IF}}<-3$.

\begin{figure}[t]
  \centering
  \includegraphics[width=0.45\textwidth]{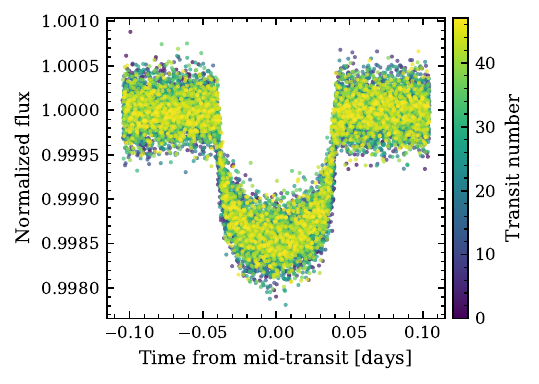}
  \caption{Folded transits from the target $L=30$ category, corresponding to $\mathrm{LLR}_{\rm theory}=900$, after the \emph{individual fits}. Each curve shows a single transit aligned by its fixed model-minimum center $t_{c,j}$. These folded light curves provide the input for the SIPVA stage.}
  \label{fig:folded_transits}
\end{figure}

\subsection{\emph{Simultaneous Impact Parameter Variation Analysis}} \label{sec:Group_fit}

{After the individual fits, we determine a fixed folding center $t_{c,j}$ for each retained transit by evaluating the transit model at the posterior-median transit parameters and locating its minimum with bounded scalar minimization. For the analysis of the \emph{Kepler} observations, this model-minimum center is distinct from the event-prediction center used to locate the transit and center the prior on $t_{{\rm e},j}$. Each transit is shifted into the common folded frame using its fixed model-minimum center $t_{c,j}$, so the folded time of cadence $i$ is $\tau_{ij}=t_{ij}-t_{c,j}$. No additional timing offset is fitted during SIPVA; the mid-transit offset in the folded frame is fixed to zero. We define $t_{\rm ref}$ as the center of the first retained transit and form $x_j=t_{c,j}-t_{\rm ref}$ in days, converting it to years before evaluating the secular impact-parameter relation. The quantities $t_{c,j}$, $t_{\rm ref}$, and $x_j$ are then held fixed throughout SIPVA. We intentionally avoid enforcing a strictly periodic ephemeris because TTVs are common in observed \emph{Kepler} systems. An example from the $L=30$ category, corresponding to $\mathrm{LLR}_{\rm theory}=900$, is shown in Figure~\ref{fig:folded_transits}.}

In the general SIPVA framework, the system-level transit parameters are shared across the retained transits of a target, while the epoch-dependent impact parameters are derived from
\[
b_j=b_0+\dot b_{\mathrm S}x_j,
\qquad
x_j=t_{c,j}-t_{\rm ref},
\]
where $x_j$ is expressed in years. In this analysis, we restrict the model to non-grazing geometries, $0\leq b_j\leq1$, for every retained transit. Because limb darkening is a stellar property, the same $q_1,q_2$ pair is used for all retained transits of a target, while remaining part of the joint posterior.

For the circular synthetic experiment, the seven-parameter sampled SIPVA vector is
\begin{equation}
\vec{\Phi}
=
\left\langle
 P,\, p,\, \rho_{\rm eff},\, q_1,\, q_2,\, b_0,\, \dot b_{\mathrm S}
\right\rangle .
\end{equation}
Here $b_0$ is the impact parameter at the reference epoch $t_{\rm ref}$, and $\dot b_{\mathrm S}$ is the SIPVA-inferred secular impact-parameter rate.
The transit-model vector used in the folded frame is
\[
\vec{\Theta}_0
=
\left\langle
0,\,P,\,p,\,\rho_{\rm eff},\,q_1,\,q_2
\right\rangle,
\]
where the leading zero fixes the folded-frame mid-transit coordinate exactly.

Consequently, the combined log-likelihood function is written
as a function of the SIPVA vector $\vec{\Phi}$ by summing the
individual-transit likelihoods after substituting
$b_j=b_0+\dot b_{\mathrm S}x_j$:
\begin{equation}
\begin{aligned}
\ell_{\rm S}(\vec{\Phi})
&= -\frac{1}{2}\sum_{j=1}^{N_T}\sum_{i=1}^{n_j}
\left[
\frac{F_{ij}-\hat{F}\!\left(\tau_{ij};\vec{\Theta}_0,
b_0+\dot b_{\mathrm S}x_j\right)}{s_j}
\right]^2 \\
&\quad -\frac{1}{2}\sum_{j=1}^{N_T}
n_j\log\!\left(2\pi s_j^2\right).
\end{aligned}
\end{equation}
In the SIPVA analysis, the effective residual scales are
estimated before MCMC rather than sampled. Let
$\mathrm{RSS}_j$ denote the residual sum of squares over all
$n_j$ retained cadences of transit $j$, evaluated at a
preliminary prior-free profile-likelihood solution obtained by
differential evolution. We define
\begin{equation}
s_{\mathrm{pool}}^2
=
\frac{\sum_{j=1}^{N_T}\mathrm{RSS}_j}
     {\sum_{j=1}^{N_T}n_j},
\qquad
s_j^2
=
\frac{\nu_0 s_{\mathrm{pool}}^2+\mathrm{RSS}_j}
     {\nu_0+n_j},
\label{eq:residual_scale_shrinkage}
\end{equation}
where $s_{\mathrm{pool}}^2$ is the pooled residual mean square.
We set the regularization constant to $\nu_0=10$.
One scalar $s_j$ is applied to every cadence of transit $j$ and
held fixed throughout the SIPVA MCMC. This partially
downweights transits with larger residual scatter while
stabilizing the variance estimates for sparsely sampled
transits, without introducing $N_T$ additional sampled noise
parameters. Consequently, uncertainty in $s_j$ is not
propagated into the SIPVA posterior. The archival per-cadence
flux uncertainties are not used in the SIPVA likelihood. The
same estimator is used for the synthetic experiment and yields
$s_j\approx\sigma_{\mathrm{phot}}$.

These residual-based noise weights are distinct from the
geometric information content of each transit. Even at fixed
$s_j$, the precision of the inferred impact parameter depends
on the number and placement of the sampled cadences and on the
sensitivity of the light curve to $b$. For Gaussian noise, the
per-transit Fisher information for $b$ is
\begin{equation}
J_j =
\frac{1}{s_j^2}
\sum_{i=1}^{n_j}
\left[
\left.
\frac{\partial \widehat{F}(\tau_{ij};
\vec{\Theta}_0,b)}
     {\partial b}
\right|_{b=b_j}
\right]^2,
\qquad
\mathrm{Var}(\widehat{b}_j)\approx J_j^{-1}.
\end{equation}
Thus, information increases with the number of informative
cadences and with the light-curve sensitivity to $b$, which is
largest near ingress and egress. The sensitivity is not uniform
in $b$; as $b\rightarrow0$, the light curve becomes less
sensitive to changes in the impact parameter.

{For the synthetic experiment only, we place a weak prior on the rate of change of the impact parameter, $\dot b_{\mathrm S}\sim\mathcal{N}\!\left(0,\left[\Delta_{\dot b}^{\mathrm{pr}}\right]^2\right)$ with $\Delta_{\dot b}^{\mathrm{pr}}=0.1\,\mathrm{yr}^{-1}$. The synthetic $b_0$ prior uses the same independently specified catalog-like $b$ prior center and width, $\Delta_{b_0}^{\mathrm{pr}}=0.2$, used for the corresponding synthetic individual fits; no individual-fit posterior output is used to construct this prior. This avoids reusing the synthetic light curves in both prior construction and likelihood evaluation. For the remaining synthetic system parameters, we retain the deliberately offset priors introduced in Sec.~\ref{sec:individual_fit_1}, which were perturbed relative to the true simulation inputs to mimic literature-based estimates. The SIPVA priors used for the \emph{Kepler} observations are independent of the individual-transit posterior distributions and are described in Sec.~\ref{sec:real_kepler}.}

Therefore, the posterior distribution of the SIPVA parameter
vector $\vec{\Phi}$ can be written as
\begin{equation}
\log P\!\left(
\vec{\Phi}
\mid
\{\tau_j,t_{c,j},F_j,s_j\}_{j=1}^{N_T}
\right)
=
\ell_{\rm S}(\vec{\Phi})
+
Q(\vec{\Phi}) .
\end{equation}
The folded time arrays $\tau_j$ and the quantities \(t_{c,j}\), \(t_{\rm ref}\), and \(x_j=t_{c,j}-t_{\rm ref}\) are fixed by the folding step, with $x_j$ expressed in years in the SIPVA model.
For the synthetic experiment, define the genuinely sampled Normal-prior set $\mathcal{N}_{\Phi}=\{P,p,\rho_{\rm eff},b_0,\dot b_{\mathrm S}\}$ and Uniform-prior set $\mathcal{U}_{\Phi}=\{q_1,q_2\}$. The log-prior contribution is
\begin{equation}\label{eq:Q_updated}
\begin{aligned}
Q(\vec{\Phi})
&= -\frac{1}{2}\sum_{x\in\mathcal{N}_{\Phi}}
\left\{
\left(\frac{x-\mu_x}{\Delta_x^{\mathrm{pr}}}\right)^2
+ \log\!\left(2\pi\left(\Delta_x^{\mathrm{pr}}\right)^2\right)
\right\}
\\
&\quad + \sum_{u\in\mathcal{U}_{\Phi}}
\log {\rm Unif}(u;a_u,b_u).
\end{aligned}
\end{equation}

{For the synthetic experiment, the posterior distribution was sampled with \texttt{emcee} using 32 walkers, initialized via differential evolution. Walkers were perturbed around the DE solution with Gaussian jitter equal to 30\% of the prior width for normally distributed parameters, while $q_1$ and $q_2$ were drawn directly from their uniform priors. For the rate-of-change parameter $\dot b_{\mathrm S}$, a narrower jitter of $0.01\,\mathrm{yr}^{-1}$ was used to improve mixing. The sampler ran for 200,000 steps with the first 25\% discarded as burn-in. Convergence was verified from trace plots and posterior histograms, which showed stationary chains and unimodal posteriors.}

Let $q_{16}$, $q_{50}$, and $q_{84}$ denote the resulting posterior percentiles of $\dot b_{\mathrm S}$. We define $\Delta\dot b_{\mathrm S}\equiv\max(q_{50}-q_{16},q_{84}-q_{50})$ and the signed, dimensionless SIPVA statistic $Z_b^{\mathrm{S}}\equiv q_{50}/\Delta\dot b_{\mathrm S}$.

\subsection{Results} \label{sec:results}

We analyzed the 300 synthetic systems described above using three detection channels: the SIPVA impact-parameter fit, the transit-by-transit impact-parameter regression, and the transit-by-transit TDV regression. Successful recovery was defined by the one-sided criteria $Z_b^{\mathrm{S}}>3$ for the SIPVA impact-parameter fit, $Z_b^{\mathrm{IF}}>3$ for the transit-by-transit impact-parameter regression, and $Z_T^{\mathrm{IF}}<-3$ for the transit-by-transit TDV regression. The TDV criterion has the opposite sign because, for the circular synthetic systems, a positive injected $\dot b$ shortens the transit and therefore produces $m_T^{\mathrm{IF}}<0$. The 250 non-null systems have $\dot b>0$ and span the fixed-model injection-strength labels $L=5,10,20,30,$ and $50$, or equivalently $\mathrm{LLR}_{\rm theory}=25,100,400,900,$ and $2500$. The 50 matched null systems have $\dot b=0$ and therefore $L=0$. Neither $L$ nor $\mathrm{LLR}_{\rm theory}$ is a fitted or profiled statistic used to declare recovery. For the ROC analysis, the continuous scores are $Z_b^{\mathrm{S}}$, $Z_b^{\mathrm{IF}}$, and $-Z_T^{\mathrm{IF}}$, respectively, so that larger values indicate stronger evidence in the injected positive-$\dot b$ direction for all three channels. The TDV statistic is sign-flipped only for this directional alignment in the ROC calculation; its raw definition remains $Z_T^{\mathrm{IF}}=m_T^{\mathrm{IF}}/\sigma_{m_T^{\mathrm{IF}}}$.

{To verify that the differences between methods are not simply driven by too–low per-transit signal-to-noise ratio (SNR), we estimate the SNR of a single transit using the boxcar approximation of \citet{Kipping_2023}:}
\begin{equation}
\label{eq:snr_per_transit}
\mathrm{SNR}_{\text {per transit }}=\frac{\delta}{\sigma_{\rm phot}} \sqrt{\frac{T_{\mathrm{dur}}}{\delta_t}},
\end{equation}
{where $\delta = p^2$ is the transit depth, and $T_{\mathrm{dur}} = \frac{P}{\pi} \arcsin \left(\frac{R_{\star}+R_p}{a}\right)$ is the transit duration for a central circular transit $(b=0)$ \citep{Seager_2003}. We include ${\Delta} t$ to account for the number of cadence samples contributing to each transit.}

For the fiducial white-noise sample, the median per-transit SNRs are 21.0,
50.5, 101.0, 105.7, and 163.4 for $L=5,10,20,30,$ and $50$,
respectively. Even the weakest category therefore lies outside the low-SNR
regime. The increase across categories arises from the category-dependent
rejection sampling required to reach each target under the
$\dot b\leq0.03~\mathrm{yr}^{-1}$ cap, which changes the realized system-parameter
distributions. It is not an intrinsic $\mathrm{LLR}_{\rm theory}$--SNR
relationship or merely a consequence of the common radius-ratio prior.

\begin{figure}[t]
\centering
\includegraphics[width=0.45\textwidth]{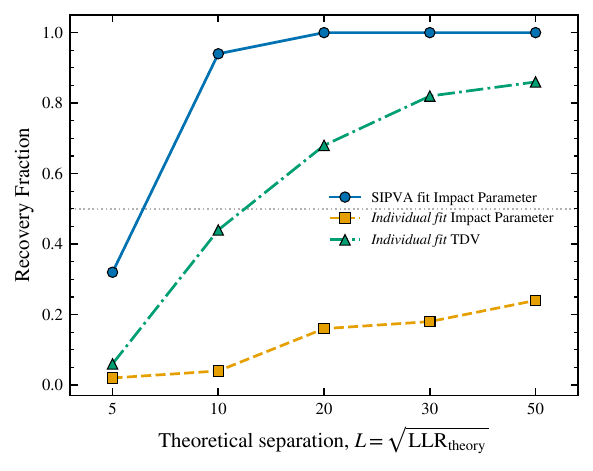}
\caption{Detection completeness for transit impact-parameter variations as a function of the fixed-model injection-strength label $L\equiv\sqrt{\mathrm{LLR}_{\rm theory}}$ in the fiducial white-noise synthetic light curves. $L$ compares the specified injected model with the specified constant-$b_0$ model and is not a fitted or profiled detection statistic. Recovery is defined by $Z_b^{\mathrm{S}}>3$, $Z_b^{\mathrm{IF}}>3$, and $Z_T^{\mathrm{IF}}<-3$ for the SIPVA impact-parameter fit, transit-by-transit impact-parameter regression, and transit-by-transit TDV regression, respectively. The negative TDV criterion follows because a positive injected $\dot b$ produces a negative duration slope in the circular synthetic systems. Each non-null category contains 50 systems at $L=5,10,20,30,$ and $50$, corresponding to $\mathrm{LLR}_{\rm theory}=25,100,400,900,$ and $2500$; the matched null set contains 50 systems with $\dot b=0$. Blue circles denote the SIPVA impact-parameter fit, orange squares the transit-by-transit impact-parameter regression, and green triangles the transit-by-transit TDV regression. The dotted horizontal line marks 50\% completeness. At these fixed thresholds, each method yielded 0/50 observed false positives in the matched null sample.}
\label{fig:tdv_recovery_curve}
\end{figure}

{Figure~\ref{fig:tdv_recovery_curve} reports recovery fractions by method across the target $L$ grid. The SIPVA impact-parameter fit has the highest completeness at every tested non-null signal strength, reaching high recovery by $L=10$ ($\mathrm{LLR}_{\rm theory}=100$). The transit-by-transit TDV regression improves with increasing $L$ but remains below the SIPVA impact-parameter fit over the tested range, while the transit-by-transit impact-parameter regression is the least complete channel. At these fixed thresholds, all three methods yielded 0/50 observed false positives in the matched null sample.
}

\begin{figure}
\centering
\includegraphics[width=0.45\textwidth]{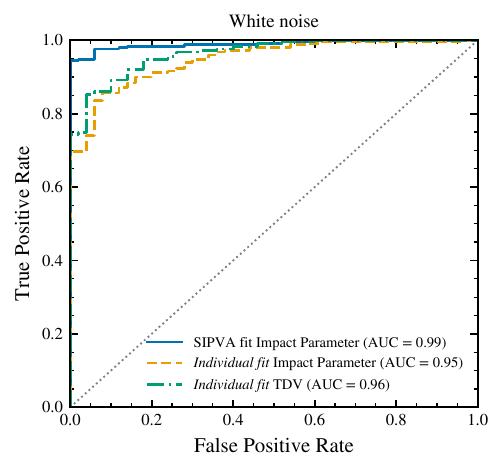}
\caption{ROC curves comparing true positive rate versus false positive rate for the SIPVA impact-parameter fit (blue line), the transit-by-transit impact-parameter regression (orange dashed line), and the transit-by-transit TDV regression (green dash-dotted line). The continuous ROC scores are $Z_b^{\mathrm{S}}$, $Z_b^{\mathrm{IF}}$, and $-Z_T^{\mathrm{IF}}$ for these three channels, respectively. The raw TDV statistic is sign-flipped only for the ROC calculation so that larger scores correspond to stronger evidence for the positive injected impact-parameter trend. The ROC sample contains 300 synthetic systems: 50 matched null systems with $\dot b=0$, treated as no signal, and 250 signal systems distributed equally across $L=5,10,20,30,$ and $50$, corresponding to $\mathrm{LLR}_{\rm theory}=25,100,400,900,$ and $2500$. The area under the curve (AUC) for each method indicates detection effectiveness: SIPVA impact-parameter fit (AUC = 0.99), transit-by-transit impact-parameter regression (AUC = 0.95), and transit-by-transit TDV regression (AUC = 0.96). The diagonal dashed line represents a random classifier.}

\label{fig:roc_curve}
\end{figure}

We evaluated threshold-dependent detection performance using ROC curves (Figure \ref{fig:roc_curve}), which plot the true positive rate against the false positive rate. For the ROC calculation, we used all 300 synthetic systems: the 50 matched null systems, treated as no-signal cases, and 250 signal systems distributed evenly across $L=5,10,20,30,$ and $50$. The area under the curve (AUC) shows that the SIPVA impact-parameter fit outperforms the transit-by-transit impact-parameter regression and the transit-by-transit TDV regression, achieving higher true-positive rates across false-positive thresholds.

This comparison isolates impact-parameter variations; in real systems, TDVs may also reflect changes in the projected mid-transit velocity (Section~\ref{sec:transit_geometry}).

\begin{figure*}
    \centering
    \begin{minipage}[c]{0.04\textwidth}
        \centering
        \rotatebox{90}{\small impact parameter $b$}
    \end{minipage}%
    \begin{minipage}[c]{0.95\textwidth}
        \centering
        \includegraphics[width=0.24\textwidth]{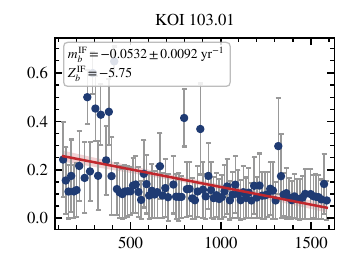}
        \includegraphics[width=0.24\textwidth]{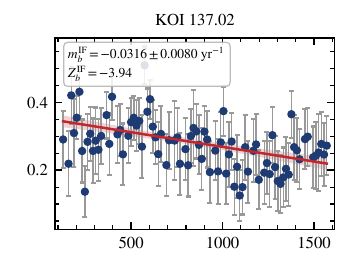}
        \includegraphics[width=0.24\textwidth]{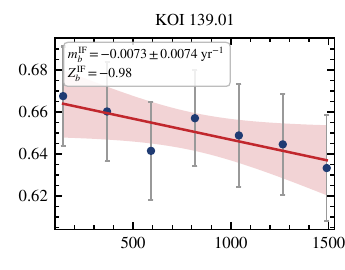}
        \includegraphics[width=0.24\textwidth]{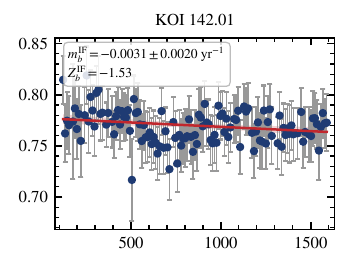}

        \includegraphics[width=0.24\textwidth]{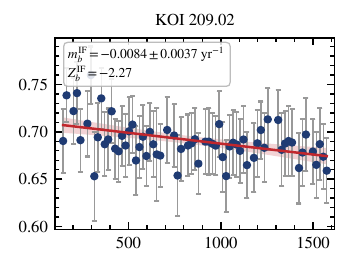}
        \includegraphics[width=0.24\textwidth]{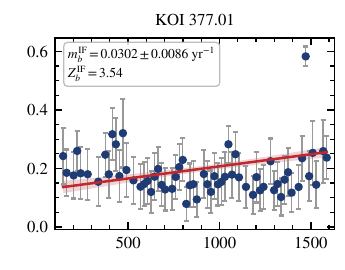}
        \includegraphics[width=0.24\textwidth]{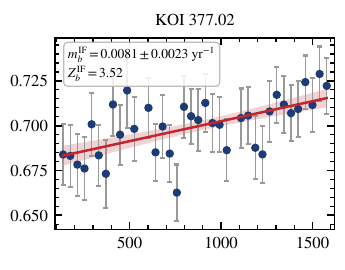}
        \includegraphics[width=0.24\textwidth]{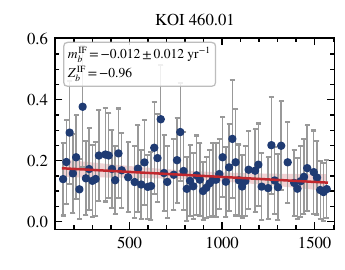}

        \includegraphics[width=0.24\textwidth]{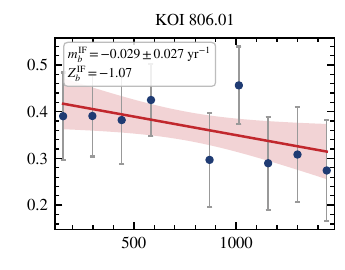}
        \includegraphics[width=0.24\textwidth]{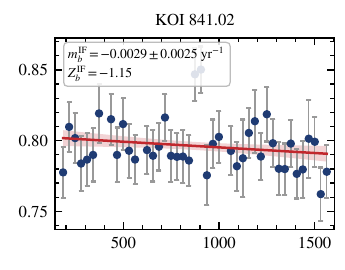}
        \includegraphics[width=0.24\textwidth]{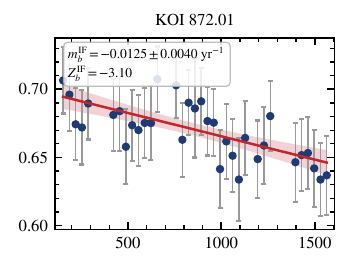}
        \includegraphics[width=0.24\textwidth]{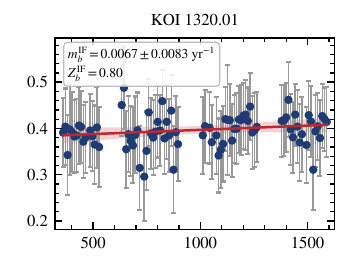}

        \includegraphics[width=0.24\textwidth]{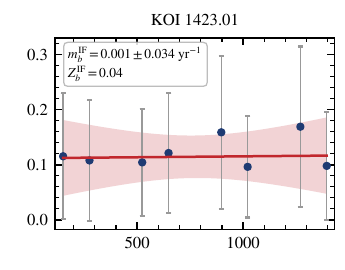}
        \includegraphics[width=0.24\textwidth]{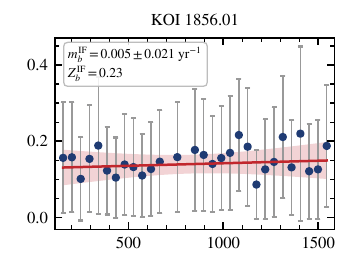}
        \includegraphics[width=0.24\textwidth]{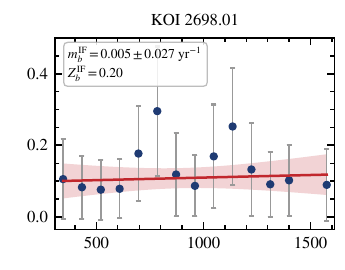}
        \includegraphics[width=0.24\textwidth]{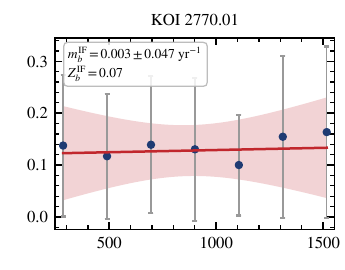}

        \vspace{0.5ex}
        {\small Fixed model-minimum center, $t_{c,j}$ [BKJD, days]}
    \end{minipage}

    \caption{Weighted linear regressions of the impact parameters inferred from the transit-by-transit fits for the 16 KOIs selected by \citet{2021AJ....162..166M}. Points show the posterior-median impact parameters ($q_{50,j}$) for retained epochs, and vertical error bars show their effective 68\% posterior uncertainties $\Delta b_j$ defined in Section~\ref{sec:individual_fit_1}. The solid red line in each panel is the weighted least-squares fit, using $w_j=(\Delta b_j)^{-2}$. The shaded region shows the pointwise ($68.3\%$) confidence band for the fitted mean relation. Each panel reports the fitted slope ($m_b^{\rm IF}$), its weighted-regression standard error $\sigma_{m_b^{\rm IF}}$, and the corresponding signed standardized slope statistic ($Z_b^{\rm IF}=m_b^{\rm IF}/\sigma_{m_b^{\rm IF}}$). The fixed model-minimum centers $t_{c,j}$ are given in Barycentric Kepler Julian Date.}
    \label{fig:impact_parameter_plots}
\end{figure*}

\section{Modeling Kepler Transits} \label{sec:real_kepler}

\citet{Holczer_2016} provided transit depth and duration measurements for 779 Kepler Objects of Interest (KOIs) observed throughout all 17 quarters of the Kepler mission. From this subset, \citet{2021AJ....162..166M} identified 16 KOIs showing long-term trends in Transit Duration Variations (TDVs). These KOIs were selected based on three criteria: (1) the absolute fitted TDV slope exceeds three times its reported uncertainty, (2) the orbital period ranges from 3 to 300 days, and the planetary radius is between 0.5 and 10 Earth radii, and (3) the host stars are part of the cleaned stellar input catalog. We applied both \emph{individual fit} (Sections \ref{sec:individual_fit_1}) and SIPVA (Section \ref{sec:Group_fit}) techniques to recover changes in the impact parameter for these 16 candidates. 

We obtained the \emph{Kepler} light curves with \texttt{lightkurve} \citep{2018ascl.soft12013L}, applied the standard Kepler quality mask, and used stellar and planetary catalog information from \texttt{kplr} \citep{2018ascl.soft07027F} and the NASA Exoplanet Archive \citep{NASAExoplanetArchive}. A TTV-aware event-prediction hierarchy is used to locate each event and define the center $t_{{\rm pred},j}$ of the individual-fit prior on $t_{{\rm e},j}$: Holczer O--C measurements are used where available, followed by a fitted-center fallback and then a linear ephemeris. These event-prediction centers are distinct from the final model-minimum folding centers $t_{c,j}$ determined after the individual fits. Predicted transits of sibling KOIs are masked before Gaussian-process detrending \citep{masuda2022pykepler}. We retain only segments with adequate sampling on both sides of the predicted center and select the available cadence only after these quality checks. Exposure durations are obtained from the \emph{Kepler} metadata. Short-cadence observations are evaluated at the exposure midpoint, whereas long-cadence observations use 15-point finite-exposure integration \citep{Kipping_2010}. This cadence treatment is used consistently in the individual-transit fits, the SIPVA likelihood, and the model evaluation used to determine $t_{c,j}$; it applies to the analysis of the \emph{Kepler} observations, not the unchanged synthetic experiment. In total, 816 candidate epochs were considered and 774 were fitted.

Asymmetric catalog uncertainties are symmetrized as $\bar{\sigma}_x=(|\sigma_{x,+}|+|\sigma_{x,-}|)/2$. We adopt $t_{{\rm e},j}\sim\mathcal{N}\!\left(t_{{\rm pred},j},(0.5\,{\rm d})^2\right)$. The Normal priors on $P$ and $b_j$ are centered on their catalog values and have standard deviations $\bar{\sigma}_P$ and $\bar{\sigma}_b$, respectively; thus, $b_j$ is re-estimated rather than fixed. The Normal prior on $p^2$ is centered on the square of the catalog planet-to-star radius ratio and has standard deviation $2p_{\rm cat}\bar{\sigma}_p$, where $p_{\rm cat}$ and $\bar{\sigma}_p$ are the catalog radius ratio and its symmetrized uncertainty. Because $\rho_{\rm eff}$ is identified with $\rho_\star$, the Normal prior on $\rho_{\rm eff}$ is centered on the catalog stellar density and uses its symmetrized catalog uncertainty, with excessively broad uncertainties capped to exclude the identified pathological low-density solutions while retaining substantially more flexibility than realistic TDV signals require. Defining $h=\sqrt e\cos\omega$ and $k=\sqrt e\sin\omega$, we assign each a Normal prior with standard deviation $0.1$. When both catalog $e$ and $\omega$ are available, the centers are $h_{\rm cat}=\sqrt{e_{\rm cat}}\cos\omega_{\rm cat}$ and $k_{\rm cat}=\sqrt{e_{\rm cat}}\sin\omega_{\rm cat}$, with $\omega_{\rm cat}$ evaluated in radians; otherwise, both priors are centered on zero. The $q_1$ and $q_2$ Normal priors are derived with \texttt{PyLDTk} \citep{Parviainen_Aigrain_2015}, and both coefficients are re-estimated for every individual transit rather than fixed. Posterior transit durations are summarized from the finite, transiting posterior samples.

The Kepler individual fit uses the transit-specific eccentric analogue of Equation~(\ref{eq:theta_circular}),
\begin{equation}\label{eq:theta_ecc_koi}
\vec{\Theta}_{{\rm ecc},j}
=
\left\langle
 t_{{\rm e},j},\, P,\, p^2,\, \rho_{\rm eff},\,
 \sqrt e\cos\omega,\, \sqrt e\sin\omega,\,
 q_1,\, q_2
\right\rangle ,
\end{equation}
The individual analysis samples $(\vec{\Theta}_{{\rm ecc},j}, b_j)$ for each transit. Only $t_{{\rm e},j}$ and $b_j$ carry explicit transit indices; the remaining components are re-estimated in each transit-specific posterior.

The Kepler SIPVA configuration uses the shared parameter vector
\begin{equation}\label{eq:phi_ecc_koi}
\begin{aligned}
\vec{\Phi}_{\rm ecc}
={}&\left\langle
 P,\, p^2,\, \rho_{\rm eff},\,
 \sqrt e\cos\omega,\,\sqrt e\sin\omega,\right.\\
&\left.
 q_1,\, q_2,\, b_0,\, \dot b_{\mathrm S}
\right\rangle .
\end{aligned}
\end{equation}
For model evaluation in the common folded frame, the corresponding fixed-timing transit vector is
\[
\vec{\Theta}_{{\rm ecc},0}
=
\left\langle
0,\,P,\,p^2,\,\rho_{\rm eff},\,
\sqrt e\cos\omega,\,\sqrt e\sin\omega,\,
q_1,\,q_2
\right\rangle,
\]
where the leading zero fixes the folded-frame mid-transit coordinate. All nine sampled components of $\vec{\Phi}_{\rm ecc}$ are shared across the retained transits of one target. The individual impact parameters are derived as
\[
b_j=b_0+\dot b_{\mathrm S}x_j,
\qquad
x_j=t_{c,j}-t_{\rm ref},
\]
where $t_{\rm ref}$ is the center of the first retained transit and the difference $t_{c,j}-t_{\rm ref}$ is converted from days to years before use. We restrict the model to non-grazing geometries, $0\leq b_j\leq1$, for every retained transit. The quantities $t_{c,j}$, $t_{\rm ref}$, and $x_j$ remain fixed during SIPVA. Each \emph{Kepler} transit is shifted into the common folded frame using its fixed $t_{c,j}$. No additional timing offset is fitted during SIPVA; the mid-transit offset in the folded frame is fixed to zero. The shared parameters $P$, $p^2$, $\rho_{\rm eff}$, $\sqrt e\cos\omega$, and $\sqrt e\sin\omega$ use the same catalog-based prior construction described above for the individual fits, while $q_1$ and $q_2$ use the same \texttt{PyLDTk}-derived Normal constraints. The remaining priors are $b_0\sim{\rm U}(0,1)$ and $\dot b_{\mathrm S}\sim\mathcal{N}\!\left(0,(0.2\,{\rm yr}^{-1})^2\right)$, with the broad numerical safety support $|\dot b_{\mathrm S}|\leq1\,{\rm yr}^{-1}$.
The SIPVA priors for the \emph{Kepler} observations are independent of the individual-transit posteriors. Individual-fit products are used only to determine the fixed folding centers, select retained epochs, initialize the numerical fit, and provide diagnostics; neither the individual $b_j$ estimates nor their weighted regression determines the $b_0$ prior. Whereas $m_b^{\rm IF}$ is obtained by regressing the independently fitted $b_j$ values, SIPVA directly infers $(b_0,\dot b_{\mathrm S})$ from the simultaneous light-curve model.

In the eccentric case, the inclination is computed from
\begingroup
\setlength{\abovedisplayskip}{6pt}
\setlength{\abovedisplayshortskip}{6pt}
\setlength{\belowdisplayskip}{10pt}
\setlength{\belowdisplayshortskip}{10pt}
\begin{equation}\label{eq:inclination_ecc_koi}
\cos i = \frac{b}{a_R}\frac{1+e\sin\omega}{1-e^2},
\end{equation}
\endgroup
which replaces the circular-orbit relation $\cos i=b/a_R$. This eccentric inclination relation is used consistently in the individual-transit model, the SIPVA likelihood, posterior-derived transit-duration calculations, and the model evaluation used to determine $t_{c,j}$. Because a primary transit light curve alone does not uniquely constrain eccentricity \citep{Kipping_2008}, the individual fits infer $\sqrt e\cos\omega$ and $\sqrt e\sin\omega$ separately for each epoch, while SIPVA uses one shared pair for all epochs of a target.

We applied both the individual fit and SIPVA methods
to these 16 candidates using the eccentric parameter
vectors defined above: $(\vec{\Theta}_{{\rm ecc},j},b_j)$
for the individual fit and $\vec{\Phi}_{\rm ecc}$ for SIPVA.
Results from the linear regression of impact parameters obtained with the \emph{individual fit} are presented in Figure~\ref{fig:impact_parameter_plots}. In each panel, the x-axis denotes the fixed model-minimum centers $t_{c,j}$ and the y-axis the corresponding independently fitted impact parameters. The signed standardized slope statistic $Z_b^{\mathrm{IF}}$ is reported in the upper-left corner, and the KOI identifier is given in the subplot title.

Our estimates are summarized in Table \ref{tab:koi_fits}. The first column lists the KOIs. Columns 2 and 3 give the individual-fit weighted-regression slope $m_b^{\mathrm{IF}}$ with its standard error and the corresponding signed standardized slope statistic $Z_b^{\mathrm{IF}}$, respectively. Columns 4 and 5 give the SIPVA posterior median and credible interval for the impact-parameter drift $\dot b_{\mathrm S}$ and its signed standardized slope statistic $Z_b^{\mathrm{S}}$, respectively. In contrast to $m_b^{\mathrm{IF}}$, which is obtained only after fitting individual $b_j$ values, $\dot b_{\mathrm S}$ is sampled directly as a parameter of the simultaneous light-curve model. Column 6 gives the SIPVA reference impact parameter $b_0$ with its posterior credible interval, and the final two columns give the orbital period and planetary radius.

\startlongtable
\begin{deluxetable*}{cccccccc}
\tabletypesize{\normalsize}
\tablecaption{\normalsize Impact-parameter summary\label{tab:koi_fits}}
\tablehead{
    \colhead{KOI} & \colhead{$m_b^{\mathrm{IF}}\pm\sigma_{m_b^{\mathrm{IF}}}$} & \colhead{$Z_b^{\mathrm{IF}}$} & \colhead{$\dot b_{\mathrm S}$} & \colhead{$Z_b^{\mathrm{S}}$} & \colhead{$b_0$} & \colhead{$P$} & \colhead{$R_p$} \\
    \colhead{} & \colhead{(yr$^{-1}$)} & \colhead{} & \colhead{(yr$^{-1}$)} & \colhead{} & \colhead{} & \colhead{(days)} & \colhead{($R_\oplus$)}
}
\startdata
103.01  & $-0.053 \pm 0.009$ &  -5.755 & $-0.126^{+0.008}_{-0.005}$ & -15.302 & $0.479^{+0.013}_{-0.013}$ &  14.911 &  2.620 \\
137.02  & $-0.032 \pm 0.008$ &  -3.945 & $-0.020^{+0.003}_{-0.003}$ &  -6.102 & $0.458^{+0.028}_{-0.031}$ &  14.859 &  5.940 \\
139.01  & $-0.007 \pm 0.007$ &  -0.982 & $-0.014^{+0.003}_{-0.003}$ &  -4.258 & $0.662^{+0.020}_{-0.021}$ & 224.779 &  7.670 \\
142.01  & $-0.003 \pm 0.002$ &  -1.530 & $-0.002^{+0.001}_{-0.001}$ &  -1.311 & $0.683^{+0.024}_{-0.025}$ &  10.916 &  3.930 \\
209.02  & $-0.008 \pm 0.004$ &  -2.265 & $-0.010^{+0.002}_{-0.003}$ &  -3.795 & $0.549^{+0.041}_{-0.045}$ &  18.796 &  7.390 \\
377.01  & $ 0.030 \pm 0.009$ &   3.536 & $ 0.002^{+0.002}_{-0.002}$ &   0.737 & $0.398^{+0.022}_{-0.025}$ &  19.271 &  7.740 \\
377.02  & $ 0.008 \pm 0.002$ &   3.523 & $ 0.010^{+0.001}_{-0.001}$ &   8.509 & $0.702^{+0.009}_{-0.009}$ &  38.908 &  7.980 \\
460.01  & $-0.012 \pm 0.012$ &  -0.960 & $-0.059^{+0.017}_{-0.018}$ &  -3.239 & $0.349^{+0.053}_{-0.054}$ &  17.588 &  4.030 \\
806.01  & $-0.029 \pm 0.027$ &  -1.073 & $-0.046^{+0.012}_{-0.015}$ &  -3.039 & $0.398^{+0.051}_{-0.055}$ & 143.206 &  8.900 \\
841.02  & $-0.003 \pm 0.003$ &  -1.145 & $-0.010^{+0.003}_{-0.003}$ &  -3.017 & $0.754^{+0.020}_{-0.024}$ &  31.330 &  6.500 \\
872.01  & $-0.012 \pm 0.004$ &  -3.096 & $-0.029^{+0.003}_{-0.003}$ &  -8.558 & $0.712^{+0.016}_{-0.017}$ &  33.601 &  8.460 \\
1320.01 & $ 0.007 \pm 0.008$ &   0.802 & $ 0.011^{+0.005}_{-0.004}$ &   2.068 & $0.410^{+0.063}_{-0.089}$ &  10.507 &  9.010 \\
1423.01 & $ 0.001 \pm 0.034$ &   0.036 & $-0.003^{+0.036}_{-0.037}$ &  -0.086 & $0.150^{+0.113}_{-0.094}$ & 124.420 & 10.300 \\
1856.01 & $ 0.005 \pm 0.021$ &   0.226 & $ 0.012^{+0.050}_{-0.046}$ &   0.248 & $0.200^{+0.149}_{-0.134}$ &  46.299 &  4.330 \\
2698.01 & $ 0.005 \pm 0.027$ &   0.198 & $-0.017^{+0.031}_{-0.031}$ &  -0.547 & $0.209^{+0.104}_{-0.124}$ &  87.972 &  3.390 \\
2770.01 & $ 0.003 \pm 0.047$ &   0.066 & $ 0.013^{+0.058}_{-0.053}$ &   0.224 & $0.156^{+0.147}_{-0.109}$ & 205.386 &  2.110 \\
\enddata
\tablecomments{$m_b^{\mathrm{IF}}$ is the slope of the per-transit individual-fit impact parameter $b$ versus the fixed model-minimum center $t_{c,j}$ from a weighted linear regression, and $\sigma_{m_b^{\mathrm{IF}}}$ is the weighted-regression standard error of that slope. The signed standardized slope statistic is $Z_b^{\mathrm{IF}}=m_b^{\mathrm{IF}}/\sigma_{m_b^{\mathrm{IF}}}$. The SIPVA impact-parameter drift $\dot b_{\mathrm S}$ is reported as the posterior median with its 16th--84th-percentile credible interval. Its signed standardized slope statistic is $Z_b^{\mathrm{S}}=q_{50}/\Delta\dot b_{\mathrm S}$, where $\Delta\dot b_{\mathrm S}$ is the larger credible half-width. The SIPVA reference-epoch impact parameter $b_0$ is reported as the posterior median with its 16th--84th-percentile credible interval. Thus, $\sigma_{m_b^{\mathrm{IF}}}$ is a frequentist weighted-regression standard error, whereas the intervals on $\dot b_{\mathrm S}$ and $b_0$ are MCMC posterior credible intervals. The orbital period $P$ and planetary radius $R_p$ are from the NASA Exoplanet Archive~\citep{NASAExoplanetArchive}. Statistics and classifications are computed from unrounded quantities. All values are rounded to three decimal places.}
\end{deluxetable*}

\begin{figure*}[p]
    \centering
    \includegraphics[width=\textwidth]{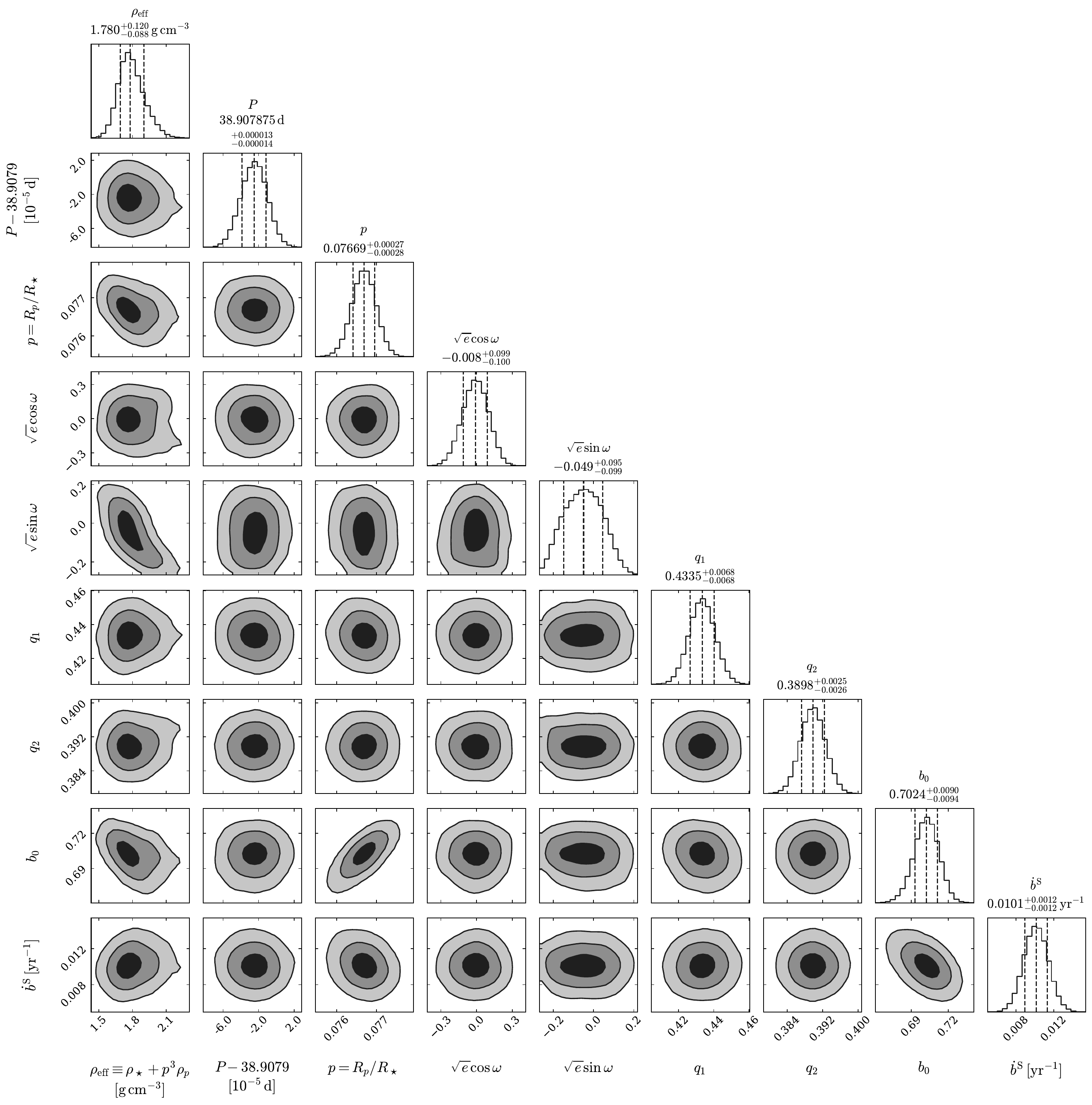}
    \caption{Posterior corner plot for the SIPVA fit to KOI 377.02, one of the nine significant SIPVA detections among the 16 candidates. The diagonal panels show the one-dimensional marginalized posterior distributions. Dashed vertical lines mark the 16th, 50th, and 84th percentiles, and each diagonal title reports the posterior median and central $68\%$ credible interval in the form ${x_{50}}^{+(x_{84}-x_{50})}_{-(x_{50}-x_{16})}$. The off-diagonal panels show the corresponding pairwise marginalized posterior distributions. The quantities, in plot order, are the fitted effective-density coordinate $\rho_{\rm eff}\,[{\rm g\,cm^{-3}}]$; the orbital period $P\,[{\rm d}]$; the planet-to-star radius ratio $p=R_p/R_\star$, displayed as the square-root transformation of the sampled $p^2$; the eccentricity variables $\sqrt e\cos\omega$ and $\sqrt e\sin\omega$; the quadratic limb-darkening coefficients $q_1$ and $q_2$; the reference-epoch impact parameter $b_0$; and the SIPVA impact-parameter slope $\dot b_{\mathrm S}\,[{\rm yr}^{-1}]$. For readability, the period axis is displayed as an offset from $38.9079\,{\rm d}$ in units of $10^{-5}\,{\rm d}$, while the diagonal title reports the full period. Quantities without listed units are dimensionless.}
    \label{fig:corner_plots}
\end{figure*}

Under the fixed $|Z_b^{\mathrm{IF}}|>3$ criterion, the transit-by-transit impact-parameter regression used here produced 5 threshold-crossing detections among the 16 planets, namely KOIs 103.01, 137.02, 377.01, 377.02, and 872.01. Under the fixed $|Z_b^{\mathrm{S}}|>3$ criterion, SIPVA produced 9 threshold-crossing detections, namely KOIs 103.01, 137.02, 139.01, 209.02, 377.02, 460.01, 806.01, 841.02, and 872.01. For KOI 377.02 (Kepler-9c), SIPVA yields $\dot b_{\mathrm S}=0.0101_{-0.0012}^{+0.0012}\:\mathrm{yr}^{-1}$, reported as the posterior median and central 68\% credible interval, comparable to the photodynamical estimate of \citet{Freudenthal_2018}; its posterior corner plot is shown in Figure~\ref{fig:corner_plots}.

Eight of the nine significant SIPVA detections have $\dot b_{\mathrm S}<0$, whereas every non-null synthetic injection has $\dot b>0$. The synthetic recovery tests therefore do not directly quantify completeness for negative slopes, particularly for trajectories approaching $b=0$, where the light curve carries less information about $b$.

\clearpage

For all planets showing significant changes in SIPVA, the direction of the impact-parameter variation aligns with the trends observed in the Transit Duration Variation fitting from \citet{Holczer_2016}. When changes in the projected mid-transit velocity are negligible, an increasing transit duration corresponds to a decreasing impact parameter.

SIPVA did not reach the fixed $|Z_b^{\mathrm{S}}|>3$ threshold for KOIs 1320.01, 1423.01, 1856.01, 2698.01, and 2770.01. For KOIs 1423.01, 1856.01, 2698.01, and 2770.01, these non-detections are plausibly due to the limited number of retained, adequately sampled transits and the large residual scatter associated with each transit. KOI 1320.01 retains a significant trend in $T_{14}$ but is not significant in the SIPVA analysis. Because transit duration depends jointly on the impact parameter and the sky-projected orbital velocity, its observed TDV cannot be uniquely attributed to a secular change in $b$. Additionally, the method did not recover a significant impact-parameter trend for KOI 142.01, as its variation is not strictly linear but rather a superposition of a wave-like function and a linear trend, as previously noted by \citet{Nesvorn__2013}. For KOI 377.01, neither the Transit Duration Variation nor SIPVA revealed a significant long-term trend.

\section{Conclusion} \label{sec:discussion}

In this paper, we developed and tested Simultaneous Impact Parameter Variation Analysis (SIPVA). The method approximates the secular evolution of the impact parameter over the \emph{Kepler} baseline as a linear trend, $b_j=b_0+\dot b_{\mathrm S}(t_{c,j}-t_{\rm ref})$, which is appropriate when the observing window samples only a small portion of the longer secular cycle \citep{2021AJ....162..166M}.

In the controlled Gaussian white-noise injections of 300 systems spanning the fixed-model injection-strength labels $L=5,10,20,30,$ and $50$ ($\mathrm{LLR}_{\rm theory}=25,100,400,900,$ and $2500$), SIPVA achieved higher recovery than the particular transit-by-transit implementation examined here. It achieved an AUC of $0.99$ and, at the fixed detection threshold, yielded 0/50 observed false positives in the matched null sample. In the selected \emph{Kepler} sample, SIPVA produced 9 threshold-crossing detections, compared with 5 from the transit-by-transit impact-parameter regression.

Taken together, these results show that SIPVA is statistically effective in the controlled injections and provides an astrophysically interpretable analysis of the selected \emph{Kepler} systems. It retains a direct connection to transit geometry and makes substantially fewer dynamical assumptions than full N-body light-curve modeling. The present formulation is best suited to systems whose secular evolution is approximately linear over the \emph{Kepler} baseline; extending SIPVA to explicitly non-linear $b(t)$ models and to joint analyses of TbVs, TDVs, and TTVs is a natural next step. Population-level inference will require explicit modeling of transit selection and detection completeness.

\section*{Data Availability}

This work is based on observations made with the NASA \emph{Kepler} spacecraft and obtained from the Mikulski Archive for Space Telescopes \citep{MASTKepler}.

This research also utilized the NASA Exoplanet Archive (\cite{NASAExoplanetArchive}), which is operated by the California Institute of Technology under contract with NASA as part of the Exoplanet Exploration Program.

All code required to reproduce the analyses and figures presented in this work is publicly available at \url{https://github.com/feynmanliu214/SIPVA}.

\begin{acknowledgments}
We thank Xian-Yu Wang for insightful discussions about modeling strategies.

\end{acknowledgments}
\vspace{1em} 

\clearpage
\bibliography{main}{}

@article{Mandel_2002,
	doi = {10.1086/345520},
  
	url = {https://doi.org/10.1086%2F345520},
  
	year = 2002,
	month = {dec},
  
	publisher = {American Astronomical Society},
  
	volume = {580},
  
	number = {2},
  
	pages = {L171--L175},
  
	author = {Kaisey Mandel and Eric Agol},
  
	title = {Analytic Light Curves for Planetary Transit Searches},
  
	journal = {The Astrophysical Journal}
}

@MISC{2018ascl.soft07027F,
       author = {{Foreman-Mackey}, Daniel},
        title = "{kplr: Tools for working with Kepler data using Python}",
 howpublished = {Astrophysics Source Code Library, record ascl:1807.027},
         year = 2018,
        month = jul,
          eid = {ascl:1807.027},
        pages = {ascl:1807.027},
archivePrefix = {ascl},
       eprint = {1807.027},
       adsurl = {https://ui.adsabs.harvard.edu/abs/2018ascl.soft07027F}
}

@ARTICLE{2021AJ....162..166M,
       author = {{Millholland}, Sarah C. and {He}, Matthias Y. and {Ford}, Eric B. and {Ragozzine}, Darin and {Fabrycky}, Daniel and {Winn}, Joshua N.},
        title = "{Evidence for a Nondichotomous Solution to the Kepler Dichotomy: Mutual Inclinations of Kepler Planetary Systems from Transit Duration Variations}",
      journal = {\aj},
         year = 2021,
        month = oct,
       volume = {162},
       number = {4},
          eid = {166},
        pages = {166},
          doi = {10.3847/1538-3881/ac0f7a},
archivePrefix = {arXiv},
       eprint = {2106.15589},
 primaryClass = {astro-ph.EP},
       adsurl = {https://ui.adsabs.harvard.edu/abs/2021AJ....162..166M}
}

@MISC{2018ascl.soft12013L,
   author = {{Lightkurve Collaboration} and {Cardoso}, J.~V.~d.~M. and
             {Hedges}, C. and {Gully-Santiago}, M. and {Saunders}, N. and
             {Cody}, A.~M. and {Barclay}, T. and {Hall}, O. and
             {Sagear}, S. and {Turtelboom}, E. and {Zhang}, J. and
             {Tzanidakis}, A. and {Mighell}, K. and {Coughlin}, J. and
             {Bell}, K. and {Berta-Thompson}, Z. and {Williams}, P. and
             {Dotson}, J. and {Barentsen}, G.},
    title = "{Lightkurve: Kepler and TESS time series analysis in Python}",
howpublished = {Astrophysics Source Code Library},
     year = 2018,
    month = dec,
archivePrefix = "ascl",
   eprint = {1812.013},
   adsurl = {http://adsabs.harvard.edu/abs/2018ascl.soft12013L},
}

@article{Shahaf_2021,
	doi = {10.1093/mnras/stab1359},
  
	url = {https://doi.org/10.1093%2Fmnras%2Fstab1359},
  
	year = 2021,
	month = {may},
  
	publisher = {Oxford University Press ({OUP})},
  
	volume = {505},
  
	number = {1},
  
	pages = {1293--1310},
  
	author = {Sahar Shahaf and Tsevi Mazeh and Shay Zucker and Daniel Fabrycky},
  
	title = {Systematic search for long-term transit duration changes in $\less$i$\greater$Kepler$\less$/i$\greater$ transiting planets},
  
	journal = {Monthly Notices of the Royal Astronomical Society}
}

@article{Judkovsky2020,
  title={Light-curve Evolution due to Secular Dynamics and the Vanishing Transits of KOI 120.01},
  author={Judkovsky, Yair and Ofir, Aviv and Aharonson, Oded},
  journal={The Astronomical Journal},
  volume={160},
  number={4},
  pages={195},
  year={2020},
  publisher={The American Astronomical Society},
  doi={10.3847/1538-3881/abb406},
}

@ARTICLE{1991AJ....102.1528W,
       author = {{Wisdom}, Jack and {Holman}, Matthew},
        title = "{Symplectic maps for the N-body problem.}",
      journal = {\aj},
         year = 1991,
        month = oct,
       volume = {102},
        pages = {1528-1538},
          doi = {10.1086/115978},
       adsurl = {https://ui.adsabs.harvard.edu/abs/1991AJ....102.1528W}
}

@ARTICLE{1999MNRAS.304..793C,
       author = {{Chambers}, J.~E.},
        title = "{A hybrid symplectic integrator that permits close encounters between massive bodies}",
      journal = {\mnras},
         year = 1999,
        month = apr,
       volume = {304},
       number = {4},
        pages = {793-799},
          doi = {10.1046/j.1365-8711.1999.02379.x},
       adsurl = {https://ui.adsabs.harvard.edu/abs/1999MNRAS.304..793C}
}

@article{Deck_2014,
   title={TTVFast: AN EFFICIENT AND ACCURATE CODE FOR TRANSIT TIMING INVERSION PROBLEMS},
   volume={787},
   ISSN={1538-4357},
   url={http://dx.doi.org/10.1088/0004-637X/787/2/132},
   DOI={10.1088/0004-637x/787/2/132},
   number={2},
   journal={The Astrophysical Journal},
   publisher={American Astronomical Society},
   author={Deck, Katherine M. and Agol, Eric and Holman, Matthew J. and Nesvorný, David},
   year={2014},
   month=may, pages={132} }

@ARTICLE{2012A&A...537A.128R,
       author = {{Rein}, H. and {Liu}, S. -F.},
        title = "{REBOUND: an open-source multi-purpose N-body code for collisional dynamics}",
      journal = {\aap},
         year = 2012,
        month = jan,
       volume = {537},
          eid = {A128},
        pages = {A128},
          doi = {10.1051/0004-6361/201118085},
archivePrefix = {arXiv},
       eprint = {1110.4876},
 primaryClass = {astro-ph.EP},
       adsurl = {https://ui.adsabs.harvard.edu/abs/2012A&A...537A.128R}
}

@ARTICLE{2015MNRAS.452..376R,
       author = {{Rein}, Hanno and {Tamayo}, Daniel},
        title = "{WHFAST: a fast and unbiased implementation of a symplectic Wisdom-Holman integrator for long-term gravitational simulations}",
      journal = {\mnras},
         year = 2015,
        month = sep,
       volume = {452},
       number = {1},
        pages = {376-388},
          doi = {10.1093/mnras/stv1257},
archivePrefix = {arXiv},
       eprint = {1506.01084},
 primaryClass = {astro-ph.EP},
       adsurl = {https://ui.adsabs.harvard.edu/abs/2015MNRAS.452..376R}
}

@ARTICLE{2005MNRAS.359..567A,
       author = {{Agol}, Eric and {Steffen}, Jason and {Sari}, Re'em and {Clarkson}, Will},
        title = "{On detecting terrestrial planets with timing of giant planet transits}",
      journal = {\mnras},
         year = 2005,
        month = may,
       volume = {359},
       number = {2},
        pages = {567-579},
          doi = {10.1111/j.1365-2966.2005.08922.x},
       adsurl = {https://ui.adsabs.harvard.edu/abs/2005MNRAS.359..567A}
}

@ARTICLE{2012ApJ...761..122L,
       author = {{Lithwick}, Yoram and {Xie}, Jiwei and {Wu}, Yanqin},
        title = "{Extracting Planet Mass and Eccentricity from TTV Data}",
      journal = {\apj},
         year = 2012,
        month = dec,
       volume = {761},
       number = {2},
          eid = {122},
        pages = {122},
          doi = {10.1088/0004-637X/761/2/122},
archivePrefix = {arXiv},
       eprint = {1207.4192},
 primaryClass = {astro-ph.EP},
       adsurl = {https://ui.adsabs.harvard.edu/abs/2012ApJ...761..122L}
}

@ARTICLE{2017AJ....154....5H,
       author = {{Hadden}, Sam and {Lithwick}, Yoram},
        title = "{Kepler Planet Masses and Eccentricities from TTV Analysis}",
      journal = {\aj},
         year = 2017,
        month = jul,
       volume = {154},
       number = {1},
          eid = {5},
        pages = {5},
          doi = {10.3847/1538-3881/aa71ef},
archivePrefix = {arXiv},
       eprint = {1611.03516},
 primaryClass = {astro-ph.EP},
       adsurl = {https://ui.adsabs.harvard.edu/abs/2017AJ....154....5H}
}

@misc{2016ascl.soft04012A,
       author = {{Agol}, Eric and {Deck}, Katherine},
        title = "{TTVFaster: First order eccentricity transit timing variations (TTVs)}",
 howpublished = {Astrophysics Source Code Library, record ascl:1604.012},
         year = 2016,
        month = apr,
          eid = {ascl:1604.012},
       adsurl = {https://ui.adsabs.harvard.edu/abs/2016ascl.soft04012A}
}

@article{Nesvorn__2010,
   title={FAST INVERSION METHOD FOR DETERMINATION OF PLANETARY PARAMETERS FROM TRANSIT TIMING VARIATIONS},
   volume={709},
   ISSN={2041-8213},
   url={http://dx.doi.org/10.1088/2041-8205/709/1/L44},
   DOI={10.1088/2041-8205/709/1/l44},
   number={1},
   journal={The Astrophysical Journal},
   publisher={American Astronomical Society},
   author={Nesvorný, David and Beaugé, Cristián},
   year={2010},
   month=jan, pages={L44–L48} }

@ARTICLE{2016ApJ...818..177A,
       author = {{Agol}, Eric and {Deck}, Katherine},
        title = "{Transit Timing to First Order in Eccentricity}",
      journal = {\apj},
         year = 2016,
        month = feb,
       volume = {818},
       number = {2},
          eid = {177},
        pages = {177},
          doi = {10.3847/0004-637X/818/2/177},
archivePrefix = {arXiv},
       eprint = {1509.01623},
 primaryClass = {astro-ph.EP},
       adsurl = {https://ui.adsabs.harvard.edu/abs/2016ApJ...818..177A}
}

@article{Parviainen_2015,
	doi = {10.1093/mnras/stv894},
  
	url = {https://doi.org/10.1093%2Fmnras%2Fstv894},
  
	year = 2015,
	month = {may},
  
	publisher = {Oxford University Press ({OUP})},
  
	volume = {450},
  
	number = {3},
  
	pages = {3233--3238},
  
	author = {Hannu Parviainen},
  
	title = {pytransit: fast and easy exoplanet transit modelling in python},
  
	journal = {Monthly Notices of the Royal Astronomical Society}
}

@article{Parviainen_Aigrain_2015,
  author = {Parviainen, Hannu and Aigrain, Suzanne},
  title = {{LDTk}: Limb Darkening Toolkit},
  journal = {Monthly Notices of the Royal Astronomical Society},
  year = {2015},
  month = nov,
  volume = {453},
  number = {4},
  pages = {3821--3826},
  doi = {10.1093/mnras/stv1857}
}

@article{Foreman_Mackey_2013,
   title={<tt>emcee</tt>: The MCMC Hammer},
   volume={125},
   ISSN={1538-3873},
   url={http://dx.doi.org/10.1086/670067},
   DOI={10.1086/670067},
   number={925},
   journal={Publications of the Astronomical Society of the Pacific},
   publisher={IOP Publishing},
   author={Foreman-Mackey, Daniel and Hogg, David W. and Lang, Dustin and Goodman, Jonathan},
   year={2013},
   month=mar, pages={306–312} }

@article{judkovsky2024kepler,
  title={Kepler Multitransiting System Physical Properties and Impact Parameter Variations},
  author={Judkovsky, Yair and Ofir, Aviv and Aharonson, Oded},
  journal={The Astronomical Journal},
  volume={167},
  number={3},
  pages={103},
  year={2024},
  publisher={The American Astronomical Society},
  doi={10.3847/1538-3881/ad16e2},
  url={https://doi.org/10.3847/1538-3881/ad16e2},
  date={2024 February 12}
}

@article{Judkovsky_2022b,
   title={Physical Properties and Impact Parameter Variations of Kepler Planets from Analytic Light-curve Modeling},
   volume={163},
   ISSN={1538-3881},
   url={http://dx.doi.org/10.3847/1538-3881/ac3d96},
   DOI={10.3847/1538-3881/ac3d96},
   number={2},
   journal={The Astronomical Journal},
   publisher={American Astronomical Society},
   author={Judkovsky, Yair and Ofir, Aviv and Aharonson, Oded},
   year={2022},
   month=jan, pages={91} }

@article{Judkovsky_2022,
   title={An Accurate 3D Analytic Model for Exoplanetary Photometry, Radial Velocity, and Astrometry},
   volume={163},
   ISSN={1538-3881},
   url={http://dx.doi.org/10.3847/1538-3881/ac3d95},
   DOI={10.3847/1538-3881/ac3d95},
   number={2},
   journal={The Astronomical Journal},
   publisher={American Astronomical Society},
   author={Judkovsky, Yair and Ofir, Aviv and Aharonson, Oded},
   year={2022},
   month=jan, pages={90} }

@article{Thompson_2018,
   title={Planetary Candidates Observed by Kepler. VIII. A Fully Automated Catalog with Measured Completeness and Reliability Based on Data Release 25},
   volume={235},
   ISSN={1538-4365},
   url={http://dx.doi.org/10.3847/1538-4365/aab4f9},
   DOI={10.3847/1538-4365/aab4f9},
   number={2},
   journal={The Astrophysical Journal Supplement Series},
   publisher={American Astronomical Society},
   author={Thompson, Susan E. and Coughlin, Jeffrey L. and Hoffman, Kelsey and Mullally, Fergal and Christiansen, Jessie L. and Burke, Christopher J. and Bryson, Steve and Batalha, Natalie and Haas, Michael R. and Catanzarite, Joseph and Rowe, Jason F. and Barentsen, Geert and Caldwell, Douglas A. and Clarke, Bruce D. and Jenkins, Jon M. and Li, Jie and Latham, David W. and Lissauer, Jack J. and Mathur, Savita and Morris, Robert L. and Seader, Shawn E. and Smith, Jeffrey C. and Klaus, Todd C. and Twicken, Joseph D. and Van Cleve, Jeffrey E. and Wohler, Bill and Akeson, Rachel and Ciardi, David R. and Cochran, William D. and Henze, Christopher E. and Howell, Steve B. and Huber, Daniel and Prša, Andrej and Ramírez, Solange V. and Morton, Timothy D. and Barclay, Thomas and Campbell, Jennifer R. and Chaplin, William J. and Charbonneau, David and Christensen-Dalsgaard, Jørgen and Dotson, Jessie L. and Doyle, Laurance and Dunham, Edward W. and Dupree, Andrea K. and Ford, Eric B. and Geary, John C. and Girouard, Forrest R. and Isaacson, Howard and Kjeldsen, Hans and Quintana, Elisa V. and Ragozzine, Darin and Shabram, Megan and Shporer, Avi and Aguirre, Victor Silva and Steffen, Jason H. and Still, Martin and Tenenbaum, Peter and Welsh, William F. and Wolfgang, Angie and Zamudio, Khadeejah A and Koch, David G. and Borucki, William J.},
   year={2018},
   month=apr, pages={38} }

@article{Holczer_2016,
   title={TRANSIT TIMING OBSERVATIONS FROM KEPLER. IX. CATALOG OF THE FULL LONG-CADENCE DATA SET},
   volume={225},
   ISSN={1538-4365},
   url={http://dx.doi.org/10.3847/0067-0049/225/1/9},
   DOI={10.3847/0067-0049/225/1/9},
   number={1},
   journal={The Astrophysical Journal Supplement Series},
   publisher={American Astronomical Society},
   author={Holczer, Tomer and Mazeh, Tsevi and Nachmani, Gil and Jontof-Hutter, Daniel and Ford, Eric B. and Fabrycky, Daniel and Ragozzine, Darin and Kane, Mackenzie and Steffen, Jason H.},
   year={2016},
   month=jul, pages={9} }

@misc{masuda2022pykepler,
  author = {Kento Masuda},
  title = {pykepler: Python Tools for Analyzing Kepler Data},
  year = {2022},
  publisher = {GitHub},
  journal = {GitHub repository},
  howpublished = {\url{https://github.com/kemasuda/pykepler}},
  license = {MIT License},
  note = {Copyright (c) 2022 Kento Masuda}
}

@article{Nesvorn__2013,
   title={KOI-142, THE KING OF TRANSIT VARIATIONS, IS A PAIR OF PLANETS NEAR THE 2:1 RESONANCE},
   volume={777},
   ISSN={1538-4357},
   url={http://dx.doi.org/10.1088/0004-637X/777/1/3},
   DOI={10.1088/0004-637x/777/1/3},
   number={1},
   journal={The Astrophysical Journal},
   publisher={American Astronomical Society},
   author={Nesvorný, David and Kipping, David and Terrell, Dirk and Hartman, Joel and Bakos, Gáspár Á. and Buchhave, Lars A.},
   year={2013},
   month=oct, pages={3} }

@article{Kipping_2008,
   title={Transit light curve analyses of eccentric exoplanets},
   volume={389},
   url={https://doi.org/10.1111/j.1365-2966.2008.13658.x},
   DOI={10.1111/j.1365-2966.2008.13658.x},
   number={3},
   journal={Monthly Notices of the Royal Astronomical Society},
   publisher={Oxford University Press (OUP)},
   author={Kipping, David M.},
   year={2008},
   month=sep, pages={1383--1390} }

@article{Kipping_2010,
   title={Binning is sinning: morphological light-curve distortions due to finite integration time: Binning is sinning},
   volume={408},
   ISSN={0035-8711},
   url={http://dx.doi.org/10.1111/j.1365-2966.2010.17242.x},
   DOI={10.1111/j.1365-2966.2010.17242.x},
   number={3},
   journal={Monthly Notices of the Royal Astronomical Society},
   publisher={Oxford University Press (OUP)},
   author={Kipping, David M.},
   year={2010},
   month=aug, pages={1758–1769} }

@inbook{Agol_2018,
   title={Transit-Timing and Duration Variations for the Discovery and Characterization of Exoplanets},
   ISBN={9783319553337},
   url={http://dx.doi.org/10.1007/978-3-319-55333-7_7},
   DOI={10.1007/978-3-319-55333-7_7},
   booktitle={Handbook of Exoplanets},
   publisher={Springer International Publishing},
   author={Agol, Eric and Fabrycky, Daniel C.},
   year={2018},
   pages={797–816} }

@article{Rein_2014,
   title={ias15: a fast, adaptive, high-order integrator for gravitational dynamics, accurate to machine precision over a billion orbits},
   volume={446},
   ISSN={0035-8711},
   url={http://dx.doi.org/10.1093/mnras/stu2164},
   DOI={10.1093/mnras/stu2164},
   number={2},
   journal={Monthly Notices of the Royal Astronomical Society},
   publisher={Oxford University Press (OUP)},
   author={Rein, Hanno and Spiegel, David S.},
   year={2014},
   month=nov, pages={1424–1437} }

@article{Rein_2019,
   title={Hybrid symplectic integrators for planetary dynamics},
   volume={485},
   ISSN={1365-2966},
   url={http://dx.doi.org/10.1093/mnras/stz769},
   DOI={10.1093/mnras/stz769},
   number={4},
   journal={Monthly Notices of the Royal Astronomical Society},
   publisher={Oxford University Press (OUP)},
   author={Rein, Hanno and Hernandez, David M and Tamayo, Daniel and Brown, Garett and Eckels, Emily and Holmes, Emma and Lau, Michelle and Leblanc, Réjean and Silburt, Ari},
   year={2019},
   month=mar, pages={5490–5497} }

@article{Rein_2019_new,
   title={High-order symplectic integrators for planetary dynamics and their implementation in rebound},
   volume={489},
   ISSN={1365-2966},
   url={http://dx.doi.org/10.1093/mnras/stz2503},
   DOI={10.1093/mnras/stz2503},
   number={4},
   journal={Monthly Notices of the Royal Astronomical Society},
   publisher={Oxford University Press (OUP)},
   author={Rein, Hanno and Tamayo, Daniel and Brown, Garett},
   year={2019},
   month=sep, pages={4632–4640} }

@article{Javaheri_2023,
   title={WHFast512: A symplectic N-body integrator for planetary systems optimized with AVX512 instructions},
   volume={6},
   ISSN={2565-6120},
   url={http://dx.doi.org/10.21105/astro.2307.05683},
   DOI={10.21105/astro.2307.05683},
   journal={The Open Journal of Astrophysics},
   publisher={Maynooth University},
   author={Javaheri, Pejvak and Rein, Hanno and Tamayo, Dan},
   year={2023},
   month=jul }

@ARTICLE{Langford2025,
      author={{Langford}, Zachary and {Agol}, Eric},
      title={A Differentiable N-body Code for Transit Timing and Dynamical Modelling---II. Photodynamics},
      journal={\mnras},
      year={2025},
      month=jun,
      volume={540},
      number={1},
      pages={106--127},
      doi={10.1093/mnras/staf687},
      archivePrefix={arXiv},
      eprint={2410.03874},
      primaryClass={astro-ph.EP}
}

@ARTICLE{Masuda2024,
       author = {{Masuda}, Kento and {Libby-Roberts}, Jessica E. and {Livingston}, John H. and {Stevenson}, Kevin B. and {Gao}, Peter and {Vissapragada}, Shreyas and {Fu}, Guangwei and {Han}, Te and {Greklek-McKeon}, Michael and {Mahadevan}, Suvrath and {Agol}, Eric and {Bello-Arufe}, Aaron and {Berta-Thompson}, Zachory and {Canas}, Caleb I. and {Chachan}, Yayaati and {Hebb}, Leslie and {Hu}, Renyu and {Kawashima}, Yui and {Knutson}, Heather A. and {Morley}, Caroline V. and {Murray}, Catriona A. and {Ohno}, Kazumasa and {Tokadjian}, Armen and {Zhang}, Xi and {Welbanks}, Luis and {Nixon}, Matthew C. and {Freedman}, Richard and {Narita}, Norio and {Fukui}, Akihiko and {de Leon}, Jerome P. and {Mori}, Mayuko and {Palle}, Enric and {Murgas}, Felipe and {Parviainen}, Hannu and {Esparza-Borges}, Emma and {Jontof-Hutter}, Daniel and {Collins}, Karen A. and {Benni}, Paul and {Barkaoui}, Khalid and {Pozuelos}, Francisco J. and {Gillon}, Michael and {Jehin}, Emmanuel and {Benkhaldoun}, Zouhair and {Hawley}, Suzanne and {Lin}, Andrea S.~J. and {Stefansson}, Gudmundur and {Bieryla}, Allyson and {Yilmaz}, Mesut and {Senavci}, Hakan Volkan and {Girardin}, Eric and {Marino}, Giuseppe and {Wang}, Gavin},
        title = "{A Fourth Planet in the Kepler-51 System Revealed by Transit Timing Variations}",
      journal = {\aj},
         year = 2024,
        month = dec,
       volume = {168},
       number = {6},
          eid = {294},
        pages = {294},
          doi = {10.3847/1538-3881/ad83d3},
archivePrefix = {arXiv},
       eprint = {2410.01625},
 primaryClass = {astro-ph.EP},
       adsurl = {https://ui.adsabs.harvard.edu/abs/2024AJ....168..294M}
}

@ARTICLE{Dai2023,
       author = {{Dai}, Fei and {Masuda}, Kento and {Beard}, Corey and {Robertson}, Paul and {Goldberg}, Max and {Batygin}, Konstantin and {Bouma}, Luke and {Lissauer}, Jack J. and {Knudstrup}, Emil and {Albrecht}, Simon and {Howard}, Andrew W. and {Knutson}, Heather A. and {Petigura}, Erik A. and {Weiss}, Lauren M. and {Isaacson}, Howard and {Kristiansen}, Martti Holst and {Osborn}, Hugh and {Wang}, Songhu and {Wang}, Xian-Yu and {Behmard}, Aida and {Greklek-McKeon}, Michael and {Vissapragada}, Shreyas and {Batalha}, Natalie M. and {Brinkman}, Casey L. and {Chontos}, Ashley and {Crossfield}, Ian and {Dressing}, Courtney and {Fetherolf}, Tara and {Fulton}, Benjamin and {Hill}, Michelle L. and {Huber}, Daniel and {Kane}, Stephen R. and {Lubin}, Jack and {MacDougall}, Mason and {Mayo}, Andrew and {Mo{\v{c}}nik}, Teo and {Akana Murphy}, Joseph M. and {Rubenzahl}, Ryan A. and {Scarsdale}, Nicholas and {Tyler}, Dakotah and {Zandt}, Judah Van and {Polanski}, Alex S. and {Schwengeler}, Hans Martin and {Terentev}, Ivan A. and {Benni}, Paul and {Bieryla}, Allyson and {Ciardi}, David and {Falk}, Ben and {Furlan}, E. and {Girardin}, Eric and {Guerra}, Pere and {Hesse}, Katharine M. and {Howell}, Steve B. and {Lillo-Box}, J. and {Matthews}, Elisabeth C. and {Twicken}, Joseph D. and {Villase{\~n}or}, Joel and {Latham}, David W. and {Jenkins}, Jon M. and {Ricker}, George R. and {Seager}, Sara and {Vanderspek}, Roland and {Winn}, Joshua N.},
        title = "{TOI-1136 is a Young, Coplanar, Aligned Planetary System in a Pristine Resonant Chain}",
      journal = {\aj},
         year = 2023,
        month = feb,
       volume = {165},
       number = {2},
          eid = {33},
        pages = {33},
          doi = {10.3847/1538-3881/aca327},
archivePrefix = {arXiv},
       eprint = {2210.09283},
 primaryClass = {astro-ph.EP},
       adsurl = {https://ui.adsabs.harvard.edu/abs/2023AJ....165...33D}
}

@ARTICLE{Grimm2018,
       author = {{Grimm}, Simon L. and {Demory}, Brice-Olivier and {Gillon}, Micha{\"e}l and {Dorn}, Caroline and {Agol}, Eric and {Burdanov}, Artem and {Delrez}, Laetitia and {Sestovic}, Marko and {Triaud}, Amaury H.~M.~J. and {Turbet}, Martin and {Bolmont}, {\'E}meline and {Caldas}, Anthony and {de Wit}, Julien and {Jehin}, Emmanu{\"e}l and {Leconte}, J{\'e}r{\'e}my and {Raymond}, Sean N. and {Van Grootel}, Val{\'e}rie and {Burgasser}, Adam J. and {Carey}, Sean and {Fabrycky}, Daniel and {Heng}, Kevin and {Hernandez}, David M. and {Ingalls}, James G. and {Lederer}, Susan and {Selsis}, Franck and {Queloz}, Didier},
        title = "{The nature of the TRAPPIST-1 exoplanets}",
      journal = {\aap},
         year = 2018,
        month = may,
       volume = {613},
          eid = {A68},
        pages = {A68},
          doi = {10.1051/0004-6361/201732233},
archivePrefix = {arXiv},
       eprint = {1802.01377},
 primaryClass = {astro-ph.EP},
       adsurl = {https://ui.adsabs.harvard.edu/abs/2018A&A...613A..68G}
}

@INPROCEEDINGS{2022BAAS...54e.360J,
       author = {{Jones}, Daniel and {Ragozzine}, Darin and {Fabrycky}, Daniel},
        title = "{PhoDyMM: the PhotoDynamical Multi-planet Model}",
    booktitle = {Bulletin of the American Astronomical Society},
         year = 2022,
       volume = {54},
        month = jun,
          eid = {102.360},
        pages = {102.360},
       adsurl = {https://ui.adsabs.harvard.edu/abs/2022BAAS...54e.360J}
}

@ARTICLE{2021ApJ...908..114Y,
       author = {{Yoffe}, Gideon and {Ofir}, Aviv and {Aharonson}, Oded},
        title = "{A Simplified Photodynamical Model for Planetary Mass Determination in Low-eccentricity Multitransiting Systems}",
      journal = {\apj},
         year = 2021,
        month = feb,
       volume = {908},
       number = {1},
          eid = {114},
        pages = {114},
          doi = {10.3847/1538-4357/abc87a},
archivePrefix = {arXiv},
       eprint = {2011.04404},
 primaryClass = {astro-ph.EP},
       adsurl = {https://ui.adsabs.harvard.edu/abs/2021ApJ...908..114Y}
}

@ARTICLE{2011Sci...333.1602D,
       author = {{Doyle}, Laurance R. and {Carter}, Joshua A. and {Fabrycky}, Daniel C. and {Slawson}, Robert W. and {Howell}, Steve B. and {Winn}, Joshua N. and {Orosz}, Jerome A. and {P{\v{r}}sa}, Andrej and {Welsh}, William F. and {Quinn}, Samuel N. and {Latham}, David and {Torres}, Guillermo and {Buchhave}, Lars A. and {Marcy}, Geoffrey W. and {Fortney}, Jonathan J. and {Shporer}, Avi and {Ford}, Eric B. and {Lissauer}, Jack J. and {Ragozzine}, Darin and {Rucker}, Michael and {Batalha}, Natalie and {Jenkins}, Jon M. and {Borucki}, William J. and {Koch}, David and {Middour}, Christopher K. and {Hall}, Jennifer R. and {McCauliff}, Sean and {Fanelli}, Michael N. and {Quintana}, Elisa V. and {Holman}, Matthew J. and {Caldwell}, Douglas A. and {Still}, Martin and {Stefanik}, Robert P. and {Brown}, Warren R. and {Esquerdo}, Gilbert A. and {Tang}, Sumin and {Furesz}, Gabor and {Geary}, John C. and {Berlind}, Perry and {Calkins}, Michael L. and {Short}, Donald R. and {Steffen}, Jason H. and {Sasselov}, Dimitar and {Dunham}, Edward W. and {Cochran}, William D. and {Boss}, Alan and {Haas}, Michael R. and {Buzasi}, Derek and {Fischer}, Debra},
        title = "{Kepler-16: A Transiting Circumbinary Planet}",
      journal = {Science},
         year = 2011,
        month = sep,
       volume = {333},
       number = {6049},
        pages = {1602},
          doi = {10.1126/science.1210923},
archivePrefix = {arXiv},
       eprint = {1109.3432},
 primaryClass = {astro-ph.EP},
       adsurl = {https://ui.adsabs.harvard.edu/abs/2011Sci...333.1602D}
}

@ARTICLE{2012Sci...337..556C,
       author = {{Carter}, Joshua A. and {Agol}, Eric and {Chaplin}, William J. and {Basu}, Sarbani and {Bedding}, Timothy R. and {Buchhave}, Lars A. and {Christensen-Dalsgaard}, J{\o}rgen and {Deck}, Katherine M. and {Elsworth}, Yvonne and {Fabrycky}, Daniel C. and {Ford}, Eric B. and {Fortney}, Jonathan J. and {Hale}, Steven J. and {Handberg}, Rasmus and {Hekker}, Saskia and {Holman}, Matthew J. and {Huber}, Daniel and {Karoff}, Christopher and {Kawaler}, Steven D. and {Kjeldsen}, Hans and {Lissauer}, Jack J. and {Lopez}, Eric D. and {Lund}, Mikkel N. and {Lundkvist}, Mia and {Metcalfe}, Travis S. and {Miglio}, Andrea and {Rogers}, Leslie A. and {Stello}, Dennis and {Borucki}, William J. and {Bryson}, Steve and {Christiansen}, Jessie L. and {Cochran}, William D. and {Geary}, John C. and {Gilliland}, Ronald L. and {Haas}, Michael R. and {Hall}, Jennifer and {Howard}, Andrew W. and {Jenkins}, Jon M. and {Klaus}, Todd and {Koch}, David G. and {Latham}, David W. and {MacQueen}, Phillip J. and {Sasselov}, Dimitar and {Steffen}, Jason H. and {Twicken}, Joseph D. and {Winn}, Joshua N.},
        title = "{Kepler-36: A Pair of Planets with Neighboring Orbits and Dissimilar Densities}",
      journal = {Science},
         year = 2012,
        month = aug,
       volume = {337},
       number = {6094},
        pages = {556},
          doi = {10.1126/science.1223269},
archivePrefix = {arXiv},
       eprint = {1206.4718},
 primaryClass = {astro-ph.EP},
       adsurl = {https://ui.adsabs.harvard.edu/abs/2012Sci...337..556C}
}

@article{Freudenthal_2018,
   title={Kepler Object of Interest Network: II. Photodynamical modelling of Kepler-9 over 8 years of transit observations},
   volume={618},
   ISSN={1432-0746},
   url={http://dx.doi.org/10.1051/0004-6361/201833436},
   DOI={10.1051/0004-6361/201833436},
   journal={Astronomy \&{} Astrophysics},
   publisher={EDP Sciences},
   author={Freudenthal, J. and von Essen, C. and Dreizler, S. and Wedemeyer, S. and Agol, E. and Morris, B. M. and Becker, A. C. and Mallonn, M. and Hoyer, S. and Ofir, A. and Tal-Or, L. and Deeg, H. J. and Herrero, E. and Ribas, I. and Khalafinejad, S. and Hernández, J. and Rodríguez S., M. M.},
   year={2018},
   month=oct,
   pages={A41},
}

@misc{MASTKepler,
  author = {{MAST}},
  title = {Kepler Mission Data},
  howpublished = {\url{https://archive.stsci.edu/kepler/}},
  doi = {10.17909/T9XG63},
  year = {2023}
}

@misc{NASAExoplanetArchive,
  author = {{NASA Exoplanet Archive}},
  title = {NASA Exoplanet Archive},
  howpublished = {\url{https://exoplanetarchive.ipac.caltech.edu}},
  year = {2023},
  doi = {10.26133/NEA1}
}

@article{Gilliland_2011,
   title={KEPLER
                    MISSION STELLAR AND INSTRUMENT NOISE PROPERTIES},
   volume={197},
   ISSN={1538-4365},
   url={http://dx.doi.org/10.1088/0067-0049/197/1/6},
   DOI={10.1088/0067-0049/197/1/6},
   number={1},
   journal={The Astrophysical Journal Supplement Series},
   publisher={American Astronomical Society},
   author={Gilliland, Ronald L. and Chaplin, William J. and Dunham, Edward W. and Argabright, Vic S. and Borucki, William J. and Basri, Gibor and Bryson, Stephen T. and Buzasi, Derek L. and Caldwell, Douglas A. and Elsworth, Yvonne P. and Jenkins, Jon M. and Koch, David G. and Kolodziejczak, Jeffrey and Miglio, Andrea and van Cleve, Jeffrey and Walkowicz, Lucianne M. and Welsh, William F.},
   year={2011},
   month=oct, pages={6} }

@article{Kipping_2023,
   title={The SNR of a transit},
   volume={523},
   ISSN={1365-2966},
   url={http://dx.doi.org/10.1093/mnras/stad1492},
   DOI={10.1093/mnras/stad1492},
   number={1},
   journal={Monthly Notices of the Royal Astronomical Society},
   publisher={Oxford University Press (OUP)},
   author={Kipping, David},
   year={2023},
   month=may, pages={1182–1191} }

@article{Seager_2003,
   title={A Unique Solution of Planet and Star Parameters from an Extrasolar Planet Transit Light Curve},
   volume={585},
   ISSN={1538-4357},
   url={http://dx.doi.org/10.1086/346105},
   DOI={10.1086/346105},
   number={2},
   journal={The Astrophysical Journal},
   publisher={American Astronomical Society},
   author={Seager, S. and Mallen‐Ornelas, G.},
   year={2003},
   month=mar, pages={1038–1055} }

@ARTICLE{Kipping2013,
       author = {{Kipping}, David M.},
        title = "{Efficient, uninformative sampling of limb darkening coefficients for two-parameter laws}",
      journal = {\mnras},
         year = 2013,
        month = nov,
       volume = {435},
       number = {3},
        pages = {2152-2160},
          doi = {10.1093/mnras/stt1435},
archivePrefix = {arXiv},
       eprint = {1308.0009},
 primaryClass = {astro-ph.SR},
       adsurl = {https://ui.adsabs.harvard.edu/abs/2013MNRAS.435.2152K}
}

@article{Miralda_Escude_2002,
   title={Orbital Perturbations of Transiting Planets: A Possible Method to Measure Stellar Quadrupoles and to Detect Earth‐Mass Planets},
   volume={564},
   ISSN={1538-4357},
   url={http://dx.doi.org/10.1086/324279},
   DOI={10.1086/324279},
   number={2},
   journal={The Astrophysical Journal},
   publisher={American Astronomical Society},
   author={Miralda‐Escude, Jordi},
   year={2002},
   month=jan, pages={1019–1023} }

@article{Boley_2020,
   title={Transit Duration Variations in Multiplanet Systems},
   volume={159},
   ISSN={1538-3881},
   url={http://dx.doi.org/10.3847/1538-3881/ab8067},
   DOI={10.3847/1538-3881/ab8067},
   number={5},
   journal={The Astronomical Journal},
   publisher={American Astronomical Society},
   author={Boley, Aaron C. and Van Laerhoven, Christa and Granados Contreras, A. P.},
   year={2020},
   month=apr, pages={207} }

@article{Szab__2012,
   title={Spin–orbit resonance, transit duration variation and possible secular perturbations in KOI-13},
   volume={421},
   ISSN={1745-3933},
   url={http://dx.doi.org/10.1111/j.1745-3933.2012.01219.x},
   DOI={10.1111/j.1745-3933.2012.01219.x},
   number={1},
   journal={Monthly Notices of the Royal Astronomical Society: Letters},
   publisher={Oxford University Press (OUP)},
   author={Szabó, Gy. M. and Pál, A. and Derekas, A. and Simon, A. E. and Szalai, T. and Kiss, L. L.},
   year={2012},
   month=mar, pages={L122–L126} }
\bibliographystyle{aasjournal}

\end{document}